\documentclass[fleqn,usenatbib]{mnras}

\usepackage{newtxtext,newtxmath}

\usepackage[T1]{fontenc}

\DeclareRobustCommand{\VAN}[3]{#2}
\let\VANthebibliography\thebibliography
\def\thebibliography{\DeclareRobustCommand{\VAN}[3]{##3}\VANthebibliography}

\usepackage{graphicx}	
\usepackage{amsmath}	

\title[Gaussian Processes for detecting AGN flares]{Using Gaussian Processes to detect AGN flares}

\author[S. A. J. McLaughlin et al.]{
Summer A. J. McLaughlin,$^{1}$\thanks{E-mail: sajmclaughlin1@sheffield.ac.uk (SAJM)}
James R. Mullaney,$^{1}$
Stuart P. Littlefair$^{1}$
\\
$^{1}$Department of Physics and Astronomy, University of Sheffield, Sheffield S3 7RH, UK
}

\date{Accepted XXX. Received YYY; in original form ZZZ}

\pubyear{2022}

\begin{document}
\label{firstpage}
\pagerange{\pageref{firstpage}--\pageref{lastpage}}
\maketitle

\begin{abstract}

\noindent A key feature of active galactic nuclei (AGN) is their variability across all wavelengths. Typically, AGN vary by a few tenths of a magnitude or more over periods lasting from hours to years. By contrast, extreme variability of AGN -- large luminosity changes that are a significant departure from the baseline variability -- are known as AGN flares. These events are rare and their timescales poorly constrained, with most of the literature focusing on individual events. It has been suggested that extreme AGN variability including flares can provide insights into the accretion processes in the disk. With surveys such as the Legacy Survey of Space and Time (LSST) promising millions of transient detections per night in the coming decade, there is a need for fast and efficient classification of AGN flares. The problem with the systematic detection of AGN flares is the requirement to detect them against a stochastically variable baseline; the ability to define a signal as a significant departure from the ever-present variability is a statistical challenge. Recently, Gaussian Processes (GPs) have revolutionised the analysis of time-series data in many areas of astronomical research. They have, however, seen limited uptake within the field of transient detection and classification. Here we investigate the efficacy of Gaussian Processes to detect AGN flares in both simulated and real optical light curves. We show that GP analysis can successfully detect AGN flares with a false-positive rate of less than seven per cent, and we present examples of AGN light curves that show extreme variability.
\end{abstract}




\section{Introduction}
\label{section:1}

Active galactic nuclei (AGN) refer to the accreting, supermassive black holes at the centres of galaxies. It is widely known that AGN vary at all wavelengths, exhibiting significant changes in luminosity on timescales of decades to minutes \citep[e.g.,][]{Ulrich1997,Graham2017,Creque-Sarbinowski2021}. Often modelled stochastically using a one-dimensional damped random walk \citep[]{Kelly2009,MacLeod2010}, the origin of this observed variability is disputed. Several models have been proposed for describing the optical variability of AGN, including accretion disk instabilities, supernovae and microlensing. Despite this debate, because reverberation mapping has shown that the broad emission lines respond to variations in the continuum emission after some time lag \citep[e.g.][]{Peterson1993}, current consensus is that continuum variations are dominated by processes intrinsic to the accretion disk, such as thermal fluctuations \citep{Kelly2009}. 

In addition to showing stochastic variability, there is evidence that AGN can exhibit extreme variability that differs significantly from the variable baseline \citep[e.g.][]{Meusinger2010,Lawrence2016,Graham2017}. These events are known as AGN flares. Current flare detections indicate that they occur over timescales of several hundreds of days \citep[]{Graham2017}, but their rarity brings up questions about how representative our existing samples are. It is thought that AGN flares are a separate phenomenon from AGN variability, though the exact cause is unknown \citep[]{Lawrence2016,Zabludoff2021}, with different studies suggesting that they could be caused by extreme instabilities in the accretion disk \citep[e.g.][]{Hawley2001}, microlensing \citep[e.g.][]{Lawrence2016,Bruce2017}, tidal disruption events \citep[]{Chan2019}, superluminous supernovae \citep[e.g.][]{Drake2011}, mergers of stellar mass black holes \citep[e.g.][]{Graham2017}, or changes in accretion state \citep[e.g.][]{Lawrence2016,MacLeod2019}. Further, magnetohydrodynamical (MHD) simulations suggest that AGN flares may be caused by energy dissipation following magnetic reconnection in the accretion disk caused by highly magnetic and turbulent processes \citep[]{Nathanail2020}.

The detection of AGN flaring events provides a window to study the accretion physics of the disk. The timescales and magnitude changes of AGN flares put constraints on MHD simulations of the accretion disk and act as important probes for AGN accretion rates over short timescales \citep[]{Lodato2010, Blagorodnova2016, Graham2017}. In addition, if follow-up spectra are acquired, reverberation mapping can enable a calculation of the size of the flaring region within the disk \citep[]{Payne2022,Zhang2013,Somalwar2022}.

Given their importance in probing the accretion physics of AGN, it is somewhat frustrating that detecting and identifying AGN flares has proven to be a challenge. In the past, they have been difficult to detect and characterise against an intrinsically variable AGN light curve \citep[]{Zabludoff2021} but improvements in imaging and machine-learning techniques are now starting to enable their detection in significant numbers \citep[]{Mattila2019}. Distinguishing a valid flare detection from the background variability presents a statistical and observational challenge \citep[]{Zabludoff2021, Gezari2021}. Even once a detection is identified, then without extensive follow-up observations, it is difficult to categorize these events with certainty (see review by \citealp{Zabludoff2021} regarding how to distinguish between types of transient). The ability to identify and distinguish between these events in real-time will be crucial to further understand not only the origins of these phenomena, but also better constrain the physics of supermassive black holes and their accretion disks, and the relationship between them. 

In the coming years, it is hoped that future high-cadence surveys will facilitate the detection and monitoring of flaring AGN in real-time \citep[]{Creque2021}. The Zwicky Transient Facility (ZTF; \citealt{Bellm2019}) and the Vera C. Rubin Observatory \citep[]{Ivezi2019} are examples of facilities that will undertake surveys to help mobilise this area of research in the coming decade with regular, high-cadence time-domain observations \citep[]{Graham2019}. These facilities will not only expand the current catalogue of AGN but will also provide insights into key unanswered questions in this field, facilitating a better understanding of AGN variability \citep[]{Creque2021}, the rates of different nuclear transient events, and what distinguishes them from each other. Real-time processing of nuclear transients detected by such surveys is critical to identifying shorter-lived, rare events and allocating follow-up resources efficiently \citep[]{Soraisam2020}.

In this era of time-domain astronomy, with facilities such as the Rubin Observatory planning to detect approximately 10 million transients per night \citep[]{Ivezi2019}, there is a vital need to be able to detect and classify AGN flares in large amounts of data, ideally before they peak to enable rapid follow-up observations. One possible means of achieving this is by using Gaussian Processes (GPs; see \S\ref{section:2}). These have emerged as a solution for modelling stochastic signals in large astronomical data-sets \citep[]{ForemanMackey2022}. The problem with searching for flares in AGN light curves is the need to detect a transient signal in data that is already intrinsically variable; there is the requirement to quantify what constitutes a significant departure from the baseline variability, and this method must be statistically robust and resistant to outliers. GPs have the potential to solve this problem, since they can be used as a means to parameterize the covariance of a data set in a statistically robust way. In this capacity, the combined use of GPs with Bayesian statistics can, in theory, determine whether a transient signal is statistically significant from a baseline model of the data.

To date there have been limited searches for AGN flares in the literature, although one notable paper is that by \citet{Graham2017}, in which 51 flare candidates were detected in a sample of over 900,000 quasar light curves from the Catalina Real-time Transient Survey. \citet{Graham2017} sought to detect flares in optical quasar light curves by first de-trending each light curve using a Theil-Sen median and then selecting contiguous sets of points (flares) above the new de-trended light curve. They subsequently used the median absolute deviation (MAD) of these flares to define a baseline level of flare activity with which to identify significant flares. In that work, they impose a minimum flare timescale by excluding flares with a duration of less than 300 days, which removes a potentially interesting region of parameter space. We propose the use of GPs to systematically detect AGN flares as a statistically robust alternative; by employing a GP to parameterize the covariance of a light curve, there is no need to {\it a priori} assume anything about the properties of the flare.

With the above in mind, the aim of this study is to assess the viability of using GPs to identify AGN flares. To do so, we first simulate the light curves of a population of variable AGN, including flaring events, then apply GP analysis to assess how successfully it identifies the latter. Next, we apply this analysis to real AGN light curves as a systematic search for flaring events. This data was obtained from the ZTF Public Data Release 6 and our sample comprises of optical, r-band light curves of Type 1 AGN \citep[]{Masci2018,Bellm2019}.

The outline of this paper is as follows: we describe the theory behind GPs (\S\ref{section:2}), the data used to investigate the efficacy of GPs (\S\ref{section:3}), the GP kernel hyper-parameter distributions (\S\ref{section:4}) and the methodology behind using GPs for the classification of AGN flares (\S\ref{section:5}) before presenting the retrieval rates of the GP analysis when dealing with different types of simulated light curves, and, finally, the light curves of real AGN (\S\ref{section:6}). We discuss our findings and future directions of studies in \S\ref{section:7}, and provide some brief concluding remarks in \S\ref{section:8}.

\section{Gaussian Processes}
\label{section:2}

Gaussian Processes (GPs) are a form of supervised machine learning, primarily used in the context of regression. A GP is often defined as a prior over functions, which generates a probability distribution over all possible functions that fit a data-set. Formally, a GP is a collection of random variables, any finite
number of which have joint Gaussian distributions. The GP is fully specified by its mean function and covariance function, which is a generalization of the Gaussian distribution \citep[]{rasmussen2006}.

Gaussian Processes are an effective non-parametric, non-linear form of regression that is powerful at handling heteroskedastic (non-uniform) uncertainties. They are regularly used in the context of astronomy for a number of tasks such as regression, modelling and classification in a variety of contexts including quasi-periodic oscillations, transient classification, AGN variability and exoplanet transits \citep[]{ForemanMackey2022}. For example, GPs have been used to de-trend variable exoplanet light curves from transit surveys \citep[]{Crossfield_2016} and also to model quasi-periodic stellar activity \citep[]{Aigrain_2012,Angus2017,Nicholson_2022}. With regards to AGN variability, GPs are commonly used to model light curves with a damped random walk kernel \citep[]{Kozlowski_2010,MacLeod2010} and indeed one of the first uses of GPs in astronomy was by \citet{Press1998} to model the variability of gravitationally-lensed quasar 0957+561 \citep[]{ForemanMackey2022}.

\begin{figure}
	\includegraphics[width=\columnwidth]{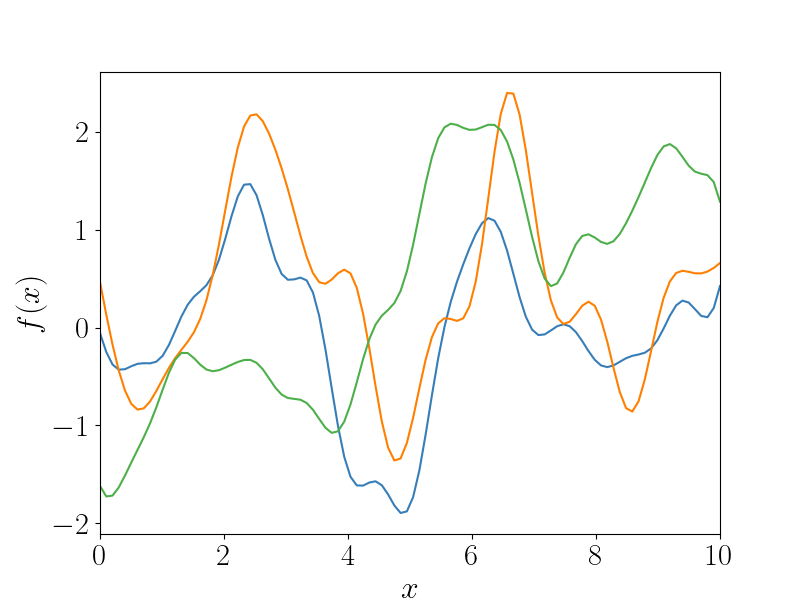}
    \caption{Three different function realisations that have been sampled from a Gaussian Process with a Matérn 3/2 kernel, using  values of 1 for both $\sigma$ and $\rho$ in Eq.~\ref{eq:gp}. Because Gaussian Processes are probabilistic in nature, the same kernel can produce different functions. Similarly, different functions (in our case light curves) can have the same, or very similar, kernel coefficients.}
    \label{fig:kernel}
\end{figure}

\begin{figure}
	\includegraphics[width=\columnwidth]{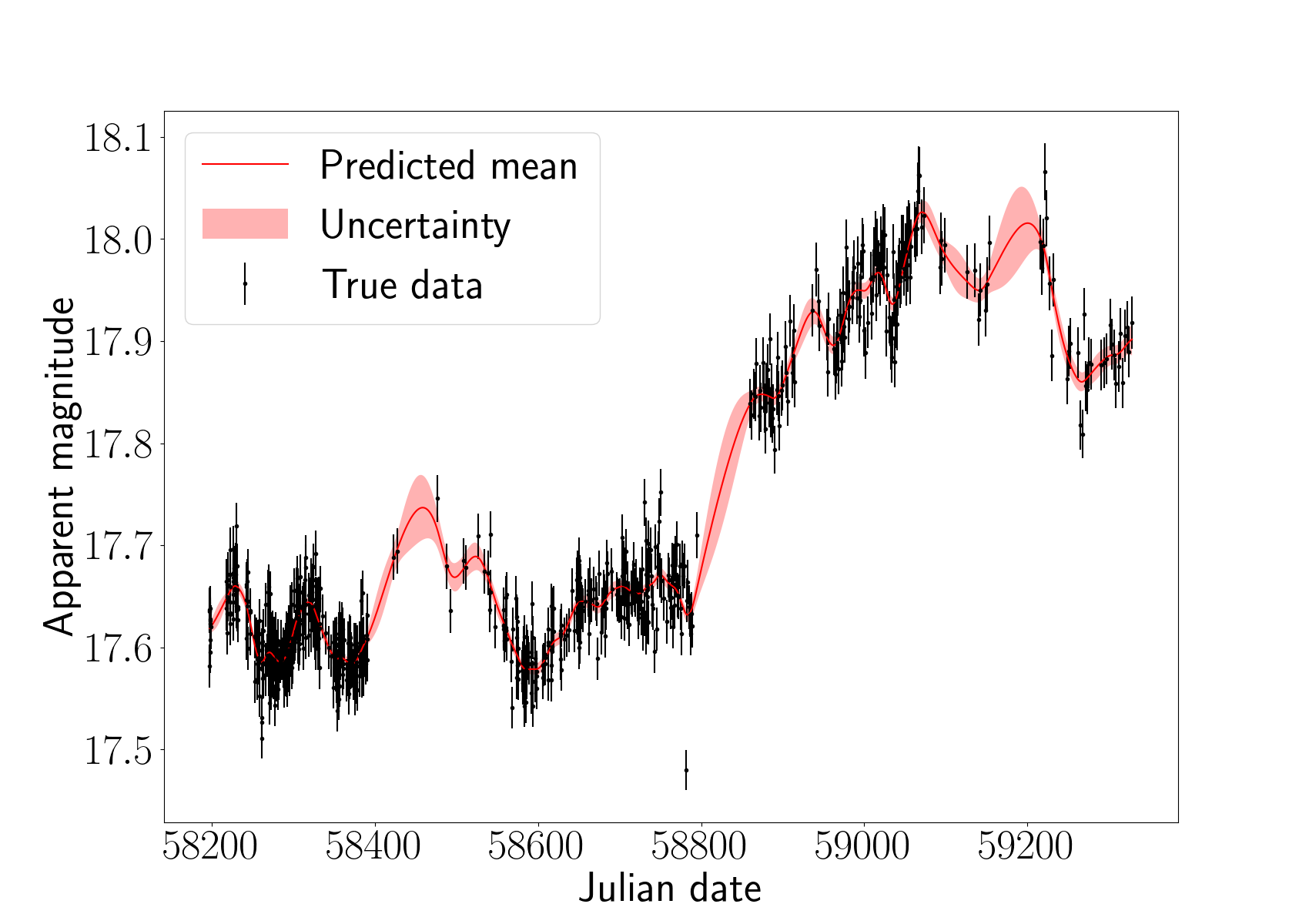}
    \caption{A Gaussian Process fit to a ZTF Type 1 AGN light curve. The red line shows the posterior mean of the Gaussian Process, given the observed data and kernel parameters. The red shaded region shows the 1-sigma uncertainty around this mean. Note how the uncertainties change depending on the density of data points in a certain region. The Gaussian Process effectively ``learns" how variable the data is, which allows it to make reasonable predictions for regions of sparse data.}
    \label{fig:gp_fit}
\end{figure}

GPs are a means of parameterising the covariance of a dataset, hence quantifying the similarity between data points. The covariance function in the context of GPs is called a kernel, and it encodes the assumptions (priors) about the underlying predictive function, for example whether it is periodic, highly variable, etc. \citep[]{rasmussen2006}. While the Gaussian Process optimises the coefficients of the kernel, it is important to choose a kernel with a functional form that is appropriate for the data in hand. In this work, we utilise the Matérn-3/2 kernel\footnote{We note that it has recently been shown that the Matérn-1/2 kernel may be more effective at reproducing AGN light curves \citep[]{Griffiths2021}, but when we repeated our analysis with this kernel there was no material difference in the results.} in which the covariance $k$ is defined as follows:

\begin{equation}
    k(\tau)=\sigma^{2}\left(1+\frac{\sqrt{3}\tau}{\rho}\right){\rm exp}\left({\frac{-\sqrt{3}\tau}{\rho}}\right),
	\label{eq:gp}
\end{equation}

\noindent
where $\tau$ is equal to the difference between all pairs of values of the independent ordinate (in this case time, i.e., $\tau=|\mathbf{t}-\mathbf{{t}'}|$, where we use bold lettering to represent a vector containing all values of time, thus ensuring that $\tau$, and hence $k(\tau)$ is a square matrix), $\sigma$ is the variability amplitude and $\rho$ is the variability timescale \citep{rasmussen2006,Foreman_Mackey_2017}. Fig.~\ref{fig:kernel} shows three different function realisations that have been sampled from a GP with a Matérn 3/2 kernel, and Fig.~\ref{fig:gp_fit} shows a GP fit to an AGN light curve with this same kernel.

GPs are widely used in the context of transient classification, but primarily used as an interpolation tool for priming sparse or noisy time series data for machine learning algorithms \citep[e.g.,][]{Villar_2020}. Within the field of astronomy research, GPs have typically been used as a pre-processing step in machine learning methods and, to our knowledge, have not been used for flare detection directly \citep{ForemanMackey2022}. We do, however, note that \cite{Graham2023} used GPs to {\it confirm} that suspected flares detected via other means do, indeed, represent significant departures from the underlying AGN variability. For flare classification directly, since the covariance function describes how all of the data points in a light curve are related to each other, it can be used as a summary statistic of the variability. This includes whether the light curve is periodic or a one-off outlier event. This therefore motivates an exploration into whether GPs can be used as a tool to classify transient astronomical events -- and specifically AGN flares -- directly.

\section{Data}
\label{section:3}

Prior to using GPs to classify real AGN light curves, we wanted to determine whether they are even a feasible means to detect flaring events. The problem with using real data for such a feasibility study is that, with AGN flares being so rare, we would need to use a large sample of AGN light curves (i.e., numbering tens to hundreds of thousands) to ensure it contains even a small handful of true flaring events. For such a large sample, however, it is unfeasible for us to know which real light curves contain true flaring events, so we cannot evaluate success rates. To overcome this, we turned to simulating light curves, which allows us to inject flares. Since we know which of the simulated light curves contain injected flares, we can determine true and false positive and negative rates. Once we have assessed the feasibility of using GPs to detect AGN flares in simulated data, we then apply it to real AGN light curves to determine whether it can, indeed, detect real AGN flaring events. We note, however, that this final step is simply an exploratory exercise; we cannot easily assess success rates on large samples of real data for the reasons outlined above. It is also important to note that while we use the simulated datasets to assess the feasibility of using GPs to identify flares, we {\it do not} use the simulated datasets to inform our priors for analysing the real ZTF light curves; instead we use the GP analysis to determine the range of typical variability parameters of each sample independently. In doing so, we ensure that any deviations from that range -- which potentially highlights the presence of a flare -- is specific to that sample.
In this section we outline how we produced our sample of simulated AGN light curves, describe how we employed GPs to analyse these simulated data, and then explain how we applied Bayesian hypothesis testing to the output of the GP analysis to calculate the probability of a light curve containing a flare.

\subsection{Simulated light curves}
\label{section:3.1}

\begin{figure}
	\includegraphics[width=\columnwidth]{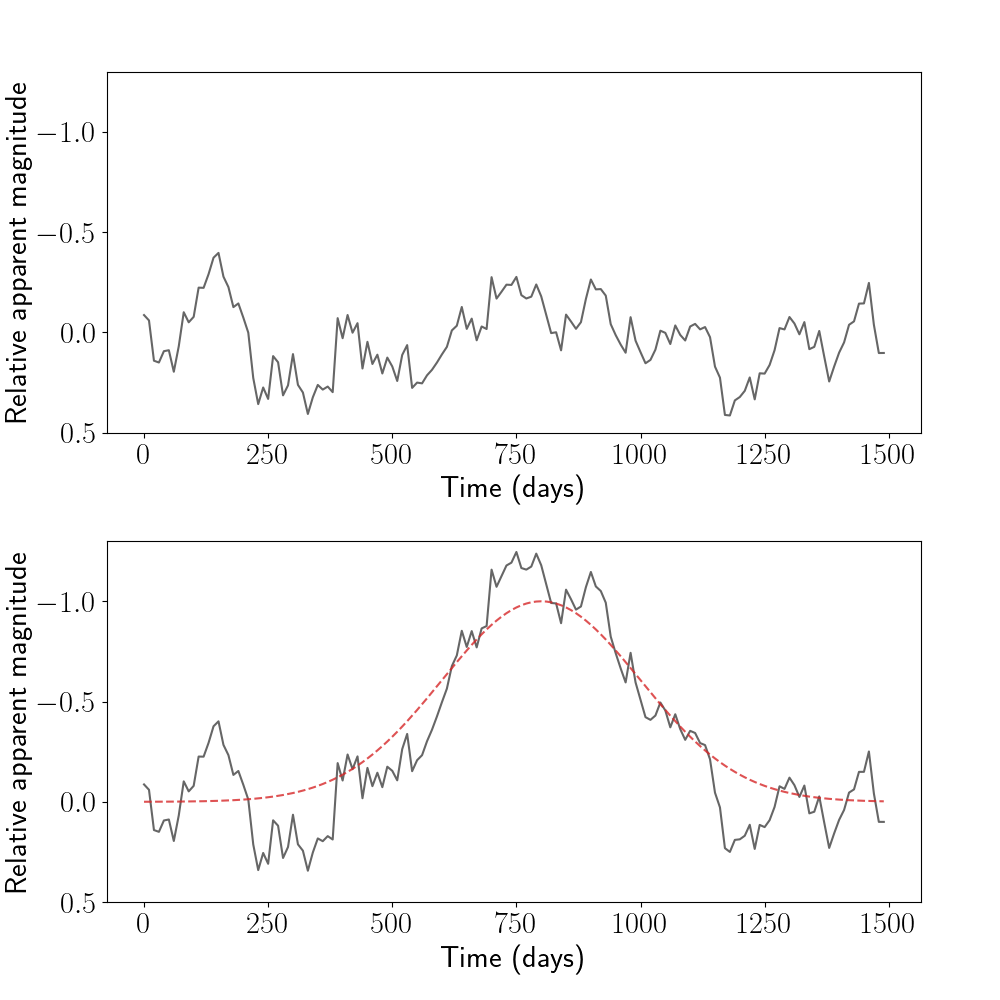}
    \caption{Top: A simulated AGN light curve using a 1D damped random walk. Bottom: The same simulated light curve with an injected 1 magnitude Gaussian flare at 800 days with a width of 200 days. The dotted red line shows this underlying Gaussian function.}
    \label{fig:Simu_lc}
\end{figure}

It has been known for over a decade that non-flaring AGN light curves are well-described by a one-dimensional damped random walk \citep[e.g.][]{Kelly2009,MacLeod2010}. This involves adding a correctional term (i.e., a damping term) to a random walk to encourage extreme deviations back to the mean value. \citet{Kelly2009} first showed that a damped random walk can statistically explain the observed light curves of AGN; they analysed 100 quasar light curves and, using a Bayesian approach, showed that this stochastic process is capable of modeling AGN light curves at an accuracy level of 0.01 - 0.02 magnitudes. 

\citet{MacLeod2010} modelled the time variability of 9000 quasars in SDSS Stripe 82 as a damped random walk and confirmed previous results \citep[e.g.][]{Kelly2009,Kozlowski_2010} that this model describes quasar light curves well. Therefore, we used this damped random walk model to simulate our own AGN light curves. We drew values of the variability parameters ($SF^{\infty}$ and $\tau$) from the distributions of best-fit variability parameters presented in \citet{MacLeod2010}. To achieve this, we randomly drew values of $\log(SF^{\infty})$ from a normal distribution with a mean of -0.8 mag and standard deviation of 0.2 mag. Then, we calculated the values of $\log(\tau)$ based on the best-fit power law in \citet{MacLeod2010}. By selecting the full range of these parameters, we intrinsically include the variability of the entire quasar population. It should also be noted that since we are assuming that AGN flares are extremely rare, we assume that the \citet{MacLeod2010} sample represents "normal" AGN variability. In their work, \citet{MacLeod2010} make use of the structure function, $ S(\tau)$, to express the long-term variability of an AGN light curve. The structure function is usually calculated as: 

\begin{equation}
    S(\tau) =\sqrt{\frac{1}{N(\tau)}\sum_{i<j}^{}[m(t_{j})-m(t_{i})]^2},
	\label{eq:1}
\end{equation}
where $\tau$ is the characteristic timescale of variability, $m(t_{i})$ is the magnitude measured at epoch $t_i$, and the sum runs over the $N(\tau)$ epochs \citep[]{Hawkins2007}.  

The structure function is computed by collecting the differences in magnitude for all points in the light curve separated by a given time lag, $\Delta$t \citep[]{MacLeod2010}. A typical AGN structure function is described by two parameters: $SF^{\infty}$ and $\tau$. The former is the difference in magnitude calculated across the longest time steps, while $\tau$ can be thought of as the damping timescale in days, upon which the value of the light curve returns to its mean. The structure function asymptotes at very large time steps (tending to $SF^{\infty}$), which corresponds to a power-law fit \citep[]{MacLeod2010}.

A light curve is generated by selecting values for $SF^{\infty}$, $\tau$ and the mean value of the light curve, $\mu$ (in our case, the values from the full distributions presented in \citet{MacLeod2010}). The magnitude $X(t)$ at a given time step $\delta t $ from a previous value $ X(t - \delta t) $ is drawn from a normal distribution with a mean and variance given by \citep[]{Kelly2009,MacLeod2010}:

\begin{equation}
    E(X(t)|X(t-\Delta t)) = \mathrm{exp}\left(-\frac{\Delta t}{\tau}\right)X(t-\Delta t) + \mu \left(1-\mathrm{exp}\left(-\frac{\Delta t}{\tau }\right)\right)
    \label{eq:4}
\end{equation}
\noindent
and
\begin{equation}    
    {\rm Var}(X(t)|X(t-\Delta t)) = 0.5(SF^{\infty})^2\left(1-\mathrm{exp}\left(-\frac{2\Delta t}{\tau }\right)\right).
	\label{eq:3}
\end{equation}

Using this approach, we simulated 10,000 AGN light curves with a cadence of 10 days and uniform uncertainties of 0.1 magnitudes. An example of one such light curve is shown in Fig.~\ref{fig:Simu_lc}. These represent our ``perfect'' simulated AGN light curves since they are regularly sampled and do not contain any anomalies (nor flares).\footnote{Hereafter, we refrain from placing quotation marks around ``perfect''.} In the following subsections we discuss the steps undertaken to modify and filter these perfect simulated light curves to include flares and to make them more representative of real, irregularly-sampled AGN light curves. To summarise, these are:

\begin{enumerate}
  \item injecting flares and simulated with a constant 10-day cadence;
  \item injecting flares and sub-sampled to match the cadence of real ZTF light curves;
  \item as (ii), but with added outliers;
  \item real ZTF with injected flares;
  \item real ZTF light curves.
\end{enumerate}

\noindent
Our objective was to investigate the ability of GPs to classify flares and non-flares in each of these cases, with each step becoming progressively more representative of observed AGN light curves.

\subsection{Light curves with injected flares}
\label{section:3.2}

To determine how GPs would handle perfect, uniformly-sampled data without outliers, we simulated 10,000 AGN light curves with a cadence of ten days. This would act as a control sample. A copy of this sample was created, and a simulated flare was injected into each light curve in the copied sample. This resulted in a control sample and a flare sample of uniformly-sampled AGN light curves, which totalled 20,000 light curves. 

The flares were simulated in two ways: (1) as Gaussian functions and (2) as gamma functions, to investigate the effect of the shape of the flare on the GP fit. Gaussian flares are symmetrical and gamma flares have a short rise-time and a longer decay. These flares were simulated with amplitudes ranging from 1-2.5 magnitudes and durations of between 100 and 1000 days. The flares were injected at random locations within each simulated light curve such that their peak lies after the first 300 days but before the last 300 days. This is to ensure that in all cases the rise and fall of the flare was included.

\subsection{Sub-sampled light curves}
\label{section:3.3}

In reality, AGN light curves are not uniformly sampled. One way to achieve non-uniformity is by randomly sub-sampling each light curve, however this would not faithfully represent real, observed light curves due to weather effects, differences between filters and large gaps in the data. For this reason, we instead interpolated the simulated light curves onto the time axis of real ZTF light curves to sub-sample them. This enabled us to investigate how GPs would handle sparsely sampled data and ensured the simulated light curves have realistic cadences. These ZTF light curves are described in \S\ref{section:3.5}.

\subsection{Light curves with added outliers}
\label{section:3.4}

In real AGN light curves, it is not uncommon to see systematic outliers in the data due to uncorrected atmospheric effects, bad pixels, etc. The GP must be robust against these effects if they are to be used as a classifying tool. Therefore, to further construct simulated light curves that were as representative of real data as possible, we added systematic outliers. To achieve this, once the light curves had been sub-sampled, we added to each light curve a contiguous pair of outliers that were five standard deviations above the variability of the individual light curve; visual inspection shows that this level of outlier is typical of ZTF light curves.

\subsection{ZTF light curves}
\label{section:3.5}

As well as using simulated data, we also used real ZTF light curves in the r band; this data was downloaded in August 2021 from Public Data Release 6 which was the most current data release at the time \citep[ZTF:][]{Masci2018,Bellm2019}. These ZTF light curves were acquired from spectroscopically-selected AGN from SDSS DR7, forming the ALPAKA catalogue \citep[]{Mullaney2013}. Of this sample, 9035 AGN are Type 1, and it is these AGN whose light curves we utilised in this paper.\footnote{We excluded Type 2 AGN from our analysis as, under the unified AGN model, we do not have a direct view of the nuclear region, meaning they are less variable and we should not -- in theory -- observe flaring events in such cases.} For the sake of a proof of concept demonstration, only the r band was considered, though it would be possible to use Gaussian Processes to perform a multi-band analysis (see \S\ref{section:7}). First, we made a copy of each AGN's ZTF light curve, and a Gaussian flare was injected into each copy. This was repeated for the injection of gamma flares. These flares were simulated as in \S\ref{section:3.2}. This created a control sample and two "flare" samples (Gaussian and gamma) of real ZTF light curves. This represents as close a sample to real AGN flaring light curves as possible, without being true flaring events.

Finally, the original sample of 9035 ZTF light curves were processed using the method outlined in the following section to determine if any of these AGN light curves would be classified as containing flares.

\section{GP kernel parameter distributions}
\label{section:4}

With our various simulated and real light curves in-hand, we next analysed them with a GP in order to calculate the optimised kernel coefficients (hereafter, hyper-parameters) of a Mat\'ern-3/2 kernel. For this, we made use of the open-source Python library \textsc{celerite} \citep{Foreman_Mackey_2017} which enables fast and scalable Gaussian Process modelling. Since \textsc{celerite} provides us with a pair of optimised hyper-parameters ($\sigma$,$\rho$) for each of our light curves, we can plot distributions of these hyper-parameters. This enables us to assess whether the distributions for flaring and non-flaring light curves reside in different regions of parameter space. If they do, then this opens up the prospect of using GP analysis to classify a light curve. In what follows, we consider the hyper-parameter distributions for each of our five different classes of light curves (i.e., those described in \S\ref{section:3}).


\subsection{Perfect light curves}
\label{section:4.1}

\begin{figure}
	\includegraphics[width=\columnwidth]{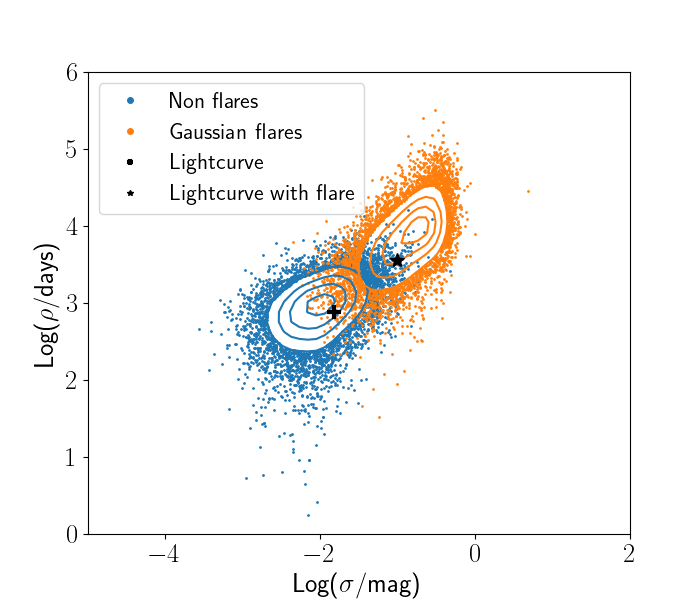}
    \caption{Distributions of flare and non-flare hyper-parameters for perfect simulated light curves with injected Gaussian flares. It is clear that the kernel hyper-parameters of light curves containing flares and light curves without flares exist in distinct but partially overlapping regions of parameter space. This demonstrates that the GP analysis finds that the covariances of these light curves are statistically different. The contours are representative of the density of the data points. The locations of the simulated light curves from Fig.~\ref{fig:Simu_lc} are shown, demonstrating that simply the injection of a flare can move a light curve's hyper-parameters from the non-flare distribution (blue points and contours) to the flare distribution (orange points and contours). The contours are representative of the density of the data points.}
    \label{fig:Gauss_hparams}
\end{figure}

\begin{figure}
	\includegraphics[width=\columnwidth]{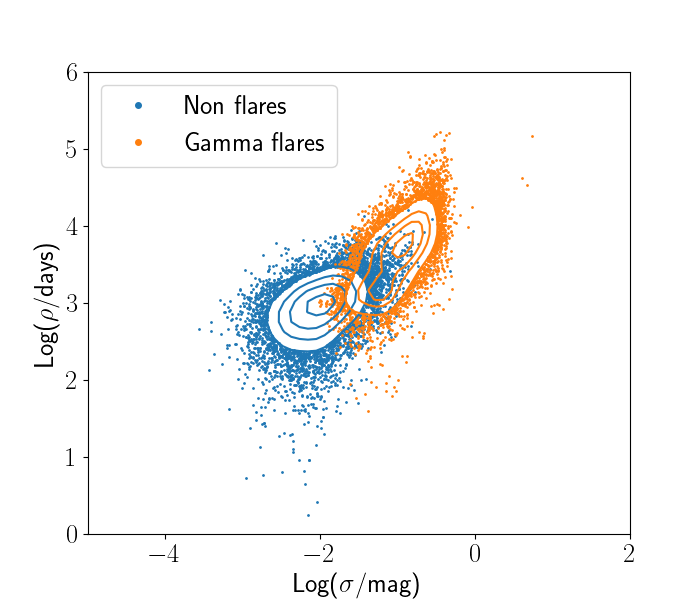}
    \caption{Distributions of flare and non-flare hyper-parameters for perfect simulated light curves with injected gamma flares. Again, it is clear that the kernel hyper-parameters of light curves containing flares and light curves without flares exist in distinct but partially overlapping regions of parameter space. The contours are representative of the density of the data points.}
    \label{fig:Gamma_hparams}
\end{figure}

The distribution of the optimised hyper-parameters for our sample of perfect light curves are shown in shown in Fig.~\ref{fig:Gauss_hparams} (for Gaussian flares) and Fig.~\ref{fig:Gamma_hparams} (for gamma flares). In these and all following plots in this section, the variability amplitude $\sigma$ increases as the variability of the light curve increases while the timescale $\rho$ increases with the timescale across which the variability is occurring.  The figures show that the hyper-parameters for flares and non-flares exist in different but partially overlapping regions of parameter space. This is the case for both Gaussian and gamma flares. This shows that the GP analysis has revealed that the covariances of these well-sampled, flaring and non-flaring light curves are statistically different. To further illustrate this, Fig.~\ref{fig:Gauss_hparams} shows the locations of the simulated light curves from Fig.~\ref{fig:Simu_lc} in hyper-parameter space, demonstrating that simply the injection of a flare into a simulated light curve moves its hyper-parameters from the non-flare distribution to the flare distribution.

\subsection{Sub-sampled light curves}
\label{section:4.2}

\begin{figure}
	\includegraphics[width=\columnwidth]{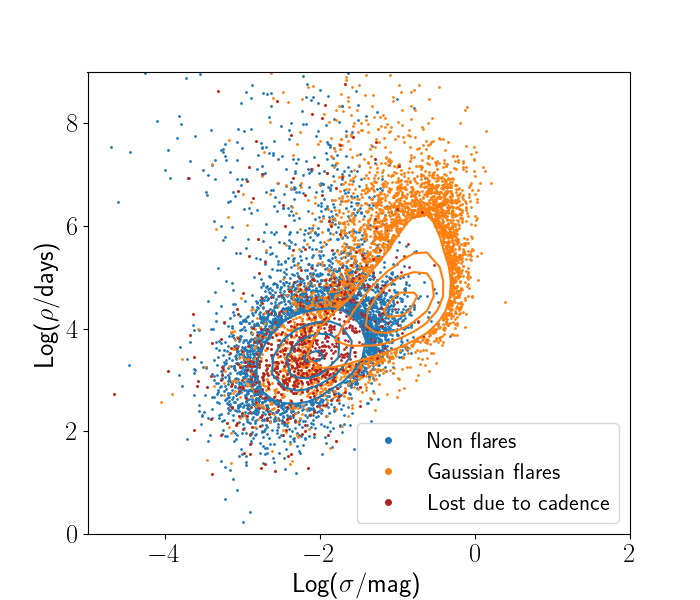}
    \caption{Distributions of flare and non-flare hyper-parameters for simulated light curves with ZTF-like cadence with Gaussian flares. Compared to the hyper-parameters of the well-sampled light curves, the distributions of light curves containing flares and light curves without flares are significantly overlapping. Light curves containing injected flares that have been significantly impacted by the post-sub-sampling cadence are shown in maroon. The contours are representative of the density of the data points.}
    \label{fig:Gauss_SS}
\end{figure}

\begin{figure}
	\includegraphics[width=\columnwidth]{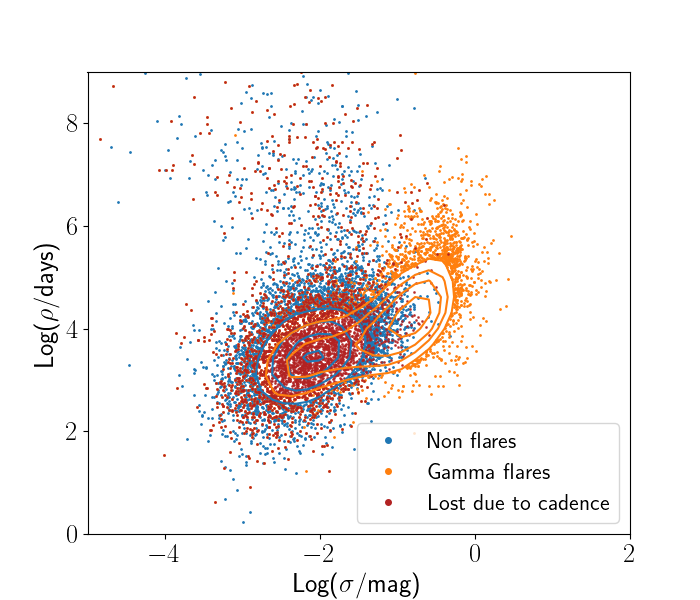}
    \caption{Distributions of flare and non-flare hyper-parameters for simulated light curves with ZTF-like cadence with gamma flares. Again, compared to the hyper-parameters of the well-sampled light curves, the distributions of light curves containing flares and light curves without flares are significantly overlapping. Light curves containing injected flares that have been significantly impacted by the post-sub-sampling cadence are shown in maroon. The contours are representative of the density of the data points.}
    \label{fig:Gamma_SS}
\end{figure}

For sub-sampled light curves, the distributions of hyper-parameters (see Figs.~\ref{fig:Gauss_SS} and \ref{fig:Gamma_SS} for Gaussian and gamma flares, respectively) overlap more than those of the perfect light curves. It should be noted, however, that this is partly due to some flares being removed by the sub-sampling, and also due to the GP analysis finding it more difficult to fit light curves with irregular cadence and gaps in the data; this is a result of the sparsity of data causing the maximum likelihood estimate of the hyper-parameters to show more scatter around the true values. In cases where the flare is largely removed by the sub-sampling, it is incorrect to regard them as false negatives, since almost all evidence of a flare has been removed from the light curve and it is important to consider it a non-flare. To determine the number of light curves in which this is the case, we summed the magnitude values of the flare points that remained post sub-sampling and ignored those light curves in which this sum was less than one. Though this choice of one is arbitrary, we investigated changing this cutoff to 0.5 and 1.5 and the results were not materially different. The use of this magnitude threshold is to ensure that light curves containing flares of which a significant proportion of the flare has been removed by the sub-sampling are treated as effectively non-flaring light curves. These poorly-sampled flares made up 11 per cent of the total number of flaring light curves, and are shown as maroon points in Fig.~\ref{fig:Gauss_SS}.

\subsection{Light curves with added outliers}
\label{section:4.3}

\begin{figure}
	\includegraphics[width=\columnwidth]{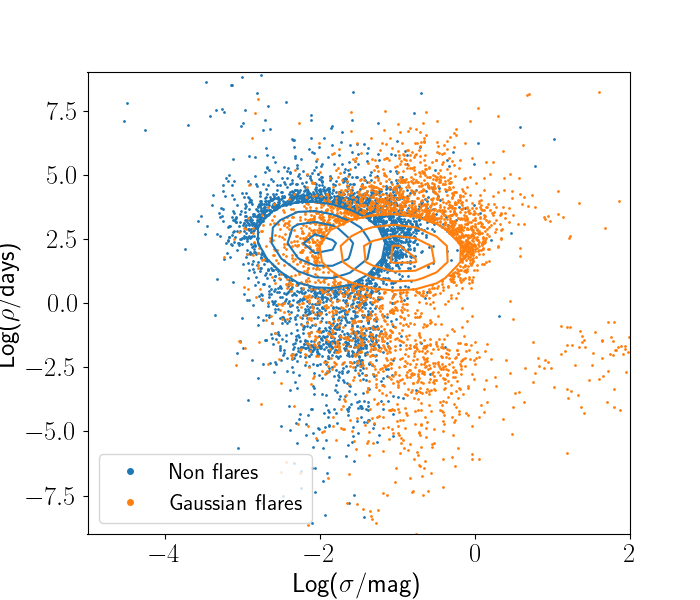}
    \caption{Distributions of flare and non-flare hyper-parameters for simulated light curves with ZTF-like cadence with added outliers and Gaussian flares. These distributions are significantly overlapping as in Fig.~\ref{fig:Gauss_SS}, and also the values of $\rho$ have been reduced. This is likely due to the injection of outliers reducing the timescale of variability calculated by the GP. The contours are representative of the density of the data points. Note that the y-axis scaling is different to the previous figures to include all of the data points.}
    \label{fig:Gauss_OL}
\end{figure}

\begin{figure}
	\includegraphics[width=\columnwidth]{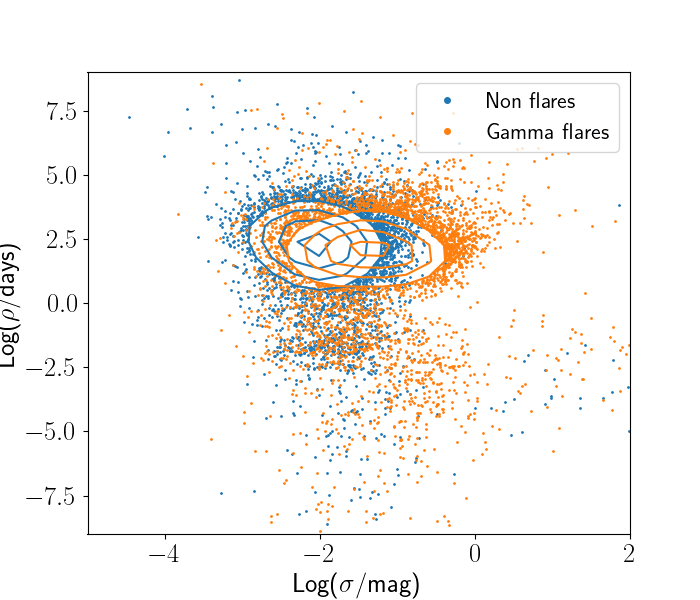}
    \caption{Distributions of flare and non-flare hyper-parameters for simulated light curves with ZTF-like cadence with added outliers and gamma flares. Again, these distributions are significantly overlapping as in Fig.~\ref{fig:Gamma_OL}, and also the values of $\rho$ have been reduced. This is likely due to the injection of outliers reducing the timescale of variability calculated by the GP. The contours are representative of the density of the data points.}
    \label{fig:Gamma_OL}
\end{figure}

The hyper-parameter distributions for the simulated, sub-sampled light curves with added outliers are shown in Figs.~\ref{fig:Gauss_OL} and \ref{fig:Gamma_OL} for the Gaussian and gamma flares respectively. There is still significant overlap between the flare and non-flare distributions, and it is clear that the flaring light curves tend to show higher values of $\sigma$ than non-flaring light curves. In both the non-flaring and flaring cases, there is a greater variance across the y-axis tending towards smaller values of $\rho$ compared to the previously discussed classes of light curve. This larger spread is a result of the addition of the outliers reducing the timescale of variability. In addition, there is greater variance across the x-axis compared to the previously-discussed classes of light curves, in that the light curves with added outliers tend towards greater values of $\sigma$. Again, this is a result of the injected outliers increasing the amplitude of variability. Finally, there are no notable differences between the Gaussian and gamma flare distributions.

\subsection{ZTF light curves with injected flares}
\label{section:4.4}

\begin{figure}
	\includegraphics[width=\columnwidth]{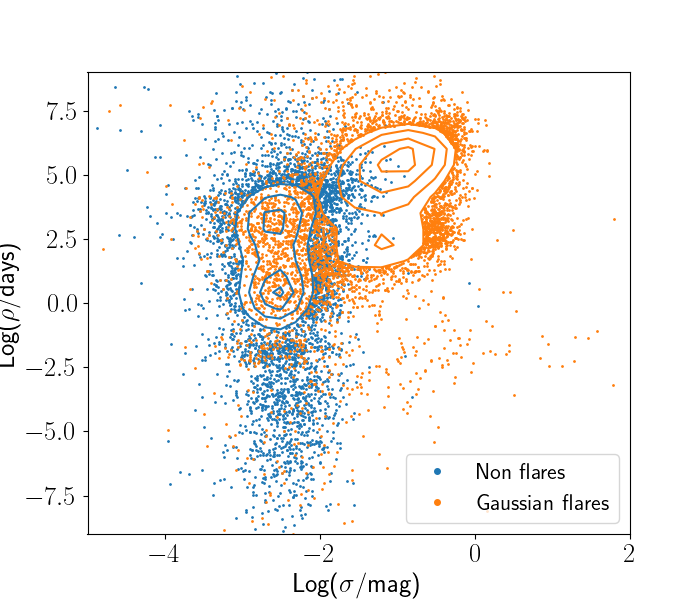}
    \caption{Distributions of hyper-parameters for ZTF light curves with injected Gaussian flares. The contours are representative of the density of the data points. Compared with Figs. \ref{fig:Gauss_hparams}, \ref{fig:Gauss_SS} and \ref{fig:Gauss_OL}, there is a much greater spread of $\rho$ values although there is still significant overlap between the distributions of light curves containing flares and light curves without flares.}
    \label{fig:ztf_inj}
\end{figure}

\begin{figure}
	\includegraphics[width=\columnwidth]{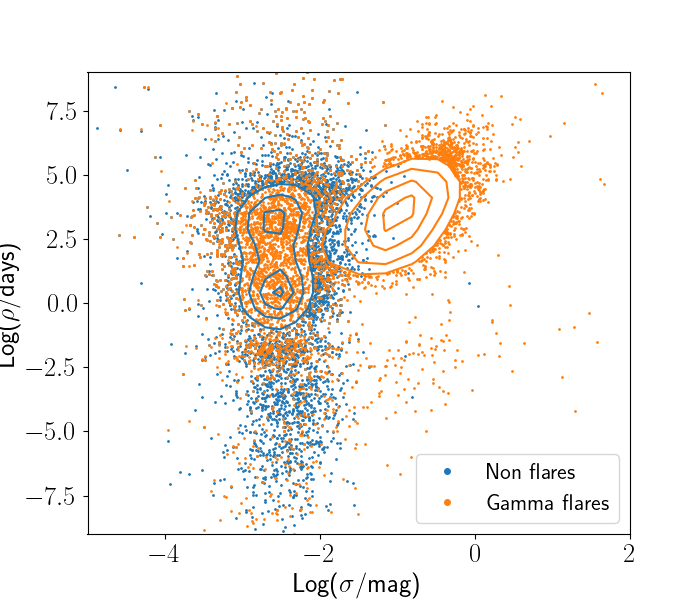}
    \caption{Distributions of hyper-parameters for ZTF light curves with injected gamma flares. The contours are representative of the density of the data points. Compared with Figs. \ref{fig:Gamma_hparams}, \ref{fig:Gamma_SS} and \ref{fig:Gamma_OL}, there is a much greater spread of $\rho$ values although there is still significant overlap between the distributions of light curves containing flares and light curves without flares.}
    \label{fig:gamma_inj}
\end{figure}

The hyper-parameter distributions for the ZTF light curves, including those with injected Gaussian and gamma flares, are shown in Figs.~\ref{fig:ztf_inj} and \ref{fig:gamma_inj}, respectively. In this case, we have assumed that the prevalence of {\it real} flares within the ZTF sample is low enough that it is reasonable to label them all as non-flaring for this part of the study. As we shall see in \S\ref{section:6.2}, it is likely that some ZTF light curves {\it are} flaring, but that their numbers are so low that (i.e., $\ll 1$ per cent) that they do not affect how we use these distributions to identify potential flares (see \S\ref{section:5}). 

The distributions of (injected) flaring and (assumed) non-flaring ZTF light curves most closely resemble those of the light curves with added outliers, displaying a larger spread across the y-axis compared to the ``perfect'' and sub-sampled light curves. However, while some overlap between hyper-parameters for flaring and non-flaring light curves is clearly present, it is somewhat less than that seen in the case of the light curves with added outliers. It is difficult to know for certain why this is the case; it may be due to the fact that we have created our simulated light curves in a way that makes them more variable than the ZTF (the median $\sigma$ value of our simulated light curves is a factor of 3 greater than that of the ZTF light curves, because the structure function values ($SF^{\infty}$ and $\tau$) used to simulate our light curves were taken from \citet{MacLeod2010} who modelled a sample of quasars rather than AGN). This means that -- all else being equal -- the injection of a flare into a simulated light curve has a smaller impact on its overall variability than the injection of the same flare into a ZTF light curve.

\section{Using GPs to identify flaring light curves}
\label{section:5}

We have demonstrated that the kernel hyper-parameters of flaring and non-flaring light curves reside in different regions of parameter space which overlap to a greater or lesser extent, depending on the class of light curve (i.e., perfect, sub-sampled, etc.). This therefore opens up the prospect of using GP analysis to identify flaring light curves. Given that the hyper-parameter distributions overlap, however, the best we can do is to assign a probability that a light curve contains a flare (i.e., $\theta=1$) or not (i.e., $\theta=0$). To achieve this, once the kernel hyper-parameters had been optimized for each light curve, we used Bayesian hypothesis testing to determine the probability of a new light curve belonging to either the flare  ($\theta=1$) or non-flare ($\theta=0$) populations. In this method, the posterior probability of a light curve containing a flare or not can be described (according to Bayes' theorem) as:

\begin{equation}
    P(\sigma, \rho, \theta | y) \propto P(y |\sigma, \rho)P(\sigma, \rho|\theta)P(\theta),
	\label{eq:bayesian}
\end{equation}
\noindent
where $\sigma$ and $\rho$ are the kernel hyper-parameters, $y$ is the data, and $P(\theta)$ is defined as a "hyper prior". In principle, $P(\theta=1)$ could be regarded as representing our {\it a priori} belief of a given light curve containing a flare, and we are thus free to choose a a value we see fit. For example, $P(\theta=1)=0.5$ would imply a prior belief that a given light curve has a 50:50 chance of containing a flare. In practice, however, we used the results of the analysis of our simulated light curves to inform us of what value of $P(\theta=1)$ gives the best compromise between numbers of false and true positives. We found, for example, that adopting $P(\theta=1)=0.001$ (which may be considered to be a reasonable estimate of the frequency of flares in AGN light curves \citep[e.g.][]{MacLeod2012,Lawrence2016}) led to a large number of simulated flares to be missed. We also investigated using $P(\theta=1)=0.5,\,0.1,\,0.01$, and found $P(\theta=1)=0.5$ resulted in a large false positive rate, while $P(\theta=1)=0.01$ suffered from a low true positive rate. Based on these results, we chose a value of $P(\theta=1)=0.1$.

We used Markov chain Monte Carlo sampling methods to sample the posterior probability distribution. We first perform an initial GP analysis of the light curve we wish to classify. This gives us the optimised values for $\sigma_c$ and $\rho_c$ for this light curve (where the subscript $c$ is used to denote the light curve we wish to classify). Next, we assign a value of 1 to $\theta_c$ if $P(\theta_c=1 | \sigma_c, \rho_c) > P(\theta_c=0 | \sigma_c, \rho_c)$ based on the distributions of $\sigma$ and $\rho$ obtained from our GP analysis, and 0 otherwise. Next, we calculate the posterior probability using:
\begin{itemize}
\item the appropriate value for $P(\theta_c)$ (i.e., 0.1 or 0.9, depending on the value of $\theta_c$);
\item the value of $P(\sigma_c, \rho_c|\theta_c)$ based on a 2D-Gaussian approximation of the hyper-parameter distributions obtained from the Gaussian Process analysis of the corresponding {\it non-flaring} light curve class.\footnote{By ``corresponding'', we mean that if we are attempting to classify a ZTF light curve then we used the hyper-parameter distribution we obtained by analysing our sample of ZTF light curves (which we assume to be dominated by non-flaring light curves), approximated using multiple 2D Gaussians.} It is important to note that we do not use the hyper-parameter distributions obtained for simulated {\it flaring} AGN as a prior for that class. Instead, we use a 2D Gaussian that encompasses a much larger region of parameter space than both the flaring and non-flaring light curves (i.e., it is a non-informative prior). This is done to ensure that we are making minimal assumptions regarding the properties of the flares since we do not know whether our simulated flares do, indeed, fully represent the true diversity of real flares.\footnote{In this regard, our analysis is agnostic to how we simulate the flares (see \S\ref{section:3.2}) since the analysis is only ascertaining whether the GP parameters of a given light curve deviate significantly from those of the non-flaring population and therefore likely to contain a flare, irrespective of its properties.};
\item the likelihood $P(y |\sigma_c, \rho_c)$ which we obtain from the Gaussian Process fit of the light curve we are classifying.
\end{itemize}

For the next step in the MCMC we randomly propose (with equal chance of choosing 0 or 1) a new value of $\theta_c$ ($=\theta_c^\prime$), and recalculate the posterior using $\theta_c^\prime$. We accept this value of $\theta_c^\prime$ with probability:
\begin{equation}
{\rm min}\left(1, \frac{P(\sigma_c, \rho_c, \theta_c^\prime | y)}{P(\sigma_c, \rho_c, \theta_c | y)}\right)
    \label{eq:accept1}
\end{equation}
(i.e., we always accept if the proposed posterior probability is greater than the current posterior probability, but accept with a probability equal to the ratio of the two posterior probabilities if $P(\sigma_c, \rho_c, \theta^\prime | y)< P(\sigma_c, \rho_c, \theta | y)$). Next, we simultaneously propose new values of $\rho_c$ and $\sigma_c$ (i.e., $\rho_c^\prime$, $\sigma_c^\prime$) and recalculate the posterior probability using these new parameters, which includes calculating $P(y | \sigma_c^\prime, \rho_c^\prime )$ using a GP. Again, we accept these values of $\sigma_c^\prime$ and $\rho_c^\prime$ with probability:
\begin{equation}
{\rm min}\left(1, \frac{P(\sigma_c^\prime, \rho_c^\prime, \theta_c | y)}{P(\sigma_c, \rho_c, \theta_c | y)}\right).
    \label{eq:accept2}
\end{equation}
Using Markov chain Monte Carlo (MCMC), we repeat the process of proposing (and, when appropriate, accepting) new $\theta$ and ($\sigma_c$, $\rho_c$) values in order to sample the posterior parameter space. We chose 12,000 steps with a burn-in of 2000 as this was sufficient for the trace to converge.

Each time we propose a new value of $\theta_c$, we add the accepted value (whether the newly-proposed value, or the old one) into a 1D array; this results in a vector of length 10,000 (excluding the burn-in) of zeroes and ones corresponding to the accepted $\theta_c$ value. The relative numbers of zeros and ones give the relative probabilities of the light curve being labelled as a flare or non-flare. As such, the final probability of the light curve containing a flare, $P_{\rm Flare}$, is thus given by the sum of this vector, divided by its length (i.e. the mean). Guided by the results from analysing our simulated data, we find that using a cutoff probability of 0.1 to define a flare gave the best compromise between true and false positives. While this may seem low, we find that most non-flaring light curves have extremely low flare probabilities.

\section{Results}
\label{section:6}

In this section we first present the retrieval rates for classifying flares and non-flares in the case of each of our classes of simulated light curves \S\ref{section:6.1}. For each class of light curve, true positive rates were calculated as the fraction of known flares with $P_{\rm Flare}$ > 0.1. Similarly, the true negative rate is the fraction of control light curves with $P_{\rm Flare}<0.1$. Afterwards, we analyse all of our unadulterated (i.e., without injected flares) ZTF light curves to see which, if any, are flagged as containing flares; the results of this ``blind'' analysis are presented in \S\ref{section:6.2}.

\subsection{Retrieval rates for simulated light curves}
\label{section:6.1}

\begin{figure}
	\includegraphics[width=\columnwidth]{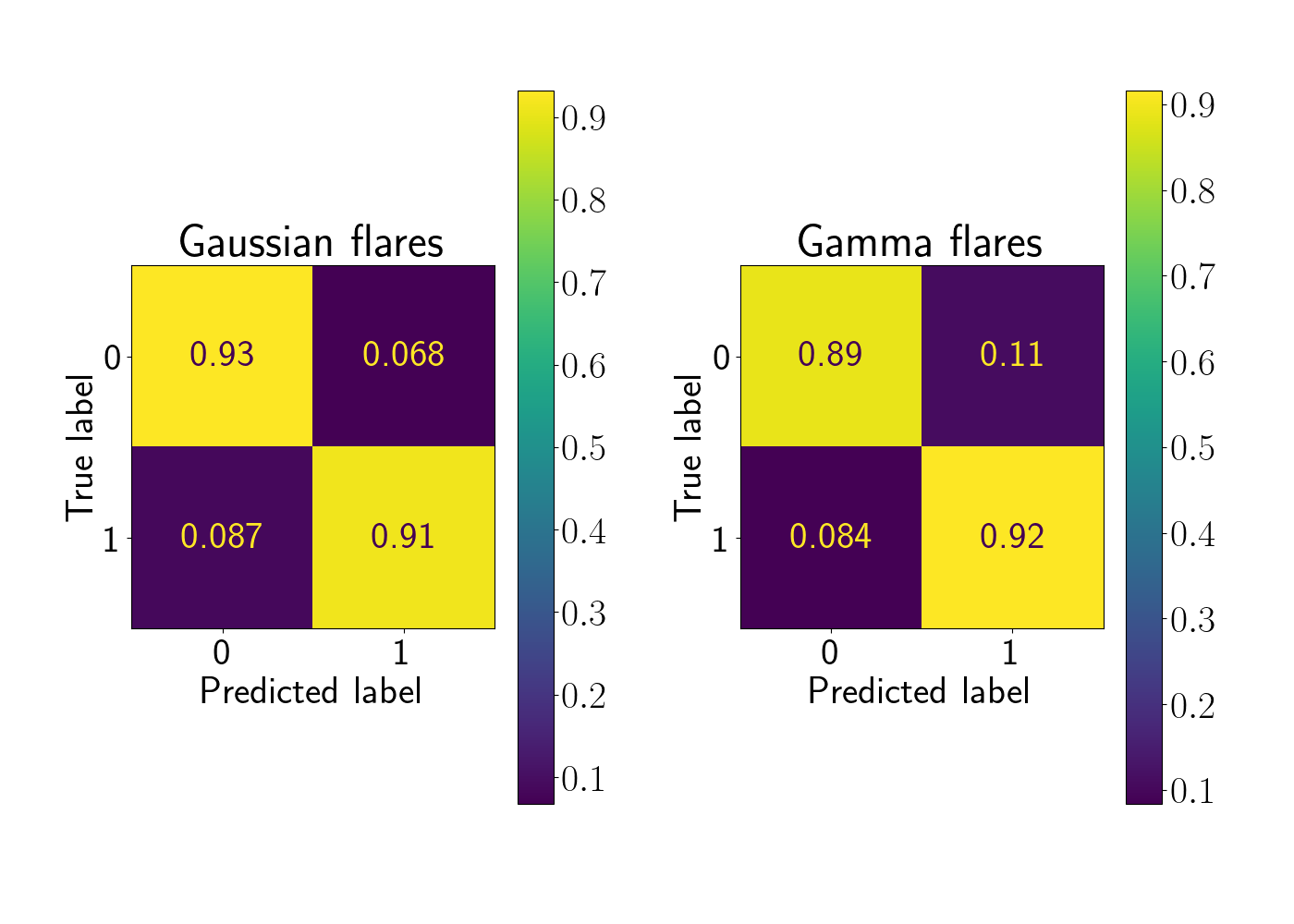}
    \caption{Confusion matrices for simulated light curves with a sampling of 10 days, in the case of injected Gaussian flares (left) and injected gamma flares (right). Note that zero and one refer to "non-flare" and "flare" respectively. The true positive rate is similar shown in the bottom right panel (91 and 92 per cent respectively), but the false positive rate shown in the top right panel is slightly higher for gamma flares (11 per cent compared to 7 per cent).}
    \label{fig:plain_cm}
\end{figure}

\begin{figure}
	\includegraphics[width=\columnwidth]{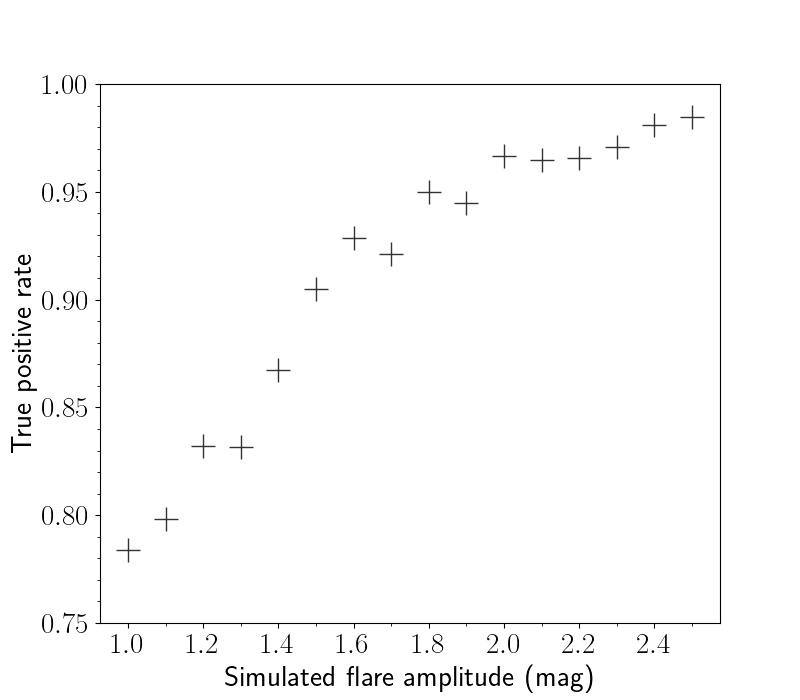}
    \caption{Retrieval rate (i.e. true positive rate) of the GP analysis as a function of simulated flare amplitude for the perfect, simulated light curves. It is clear that the retrieval rate of the GP analysis decreases as the simulated flare amplitude decreases, since the hyper-parameters of these light curves are more likely to reside in the overlapping region between flares and non-flares.}
    \label{fig:retrieval}
\end{figure}

The confusion matrices for our perfect simulated light curves with injected Gaussian and gamma flares are shown in Fig.~\ref{fig:plain_cm}. The true positive rates are similar for both types of flare (91 and 92 per cent respectively), but the false positive rate is slightly higher for gamma flares (11 per cent compared to 7 per cent). Though the simulated flare parameters must be selected arbitrarily due to the rarity of AGN flares, we investigated the change in retrieval rates of simulated flares with specific properties. We found that the retrieval rates of the GP analysis decrease as the duration of the flare increases, and the amplitude of the flare decreases. For example, 95 per cent of flares with magnitude greater than 1.5 and 99 per cent of flares with duration less than 500 days are successfully detected by the GP analysis. Fig.\ref{fig:retrieval} demonstrates the retrieval rate as a function of simulated flare amplitude, showing that the lowest amplitude flares are most difficult to detect by the GP analysis. The GP analysis is clearly struggling to distinguish simulated flares with an amplitude of one magnitude or less from the underlying variability.

\begin{figure}
	\includegraphics[width=\columnwidth]{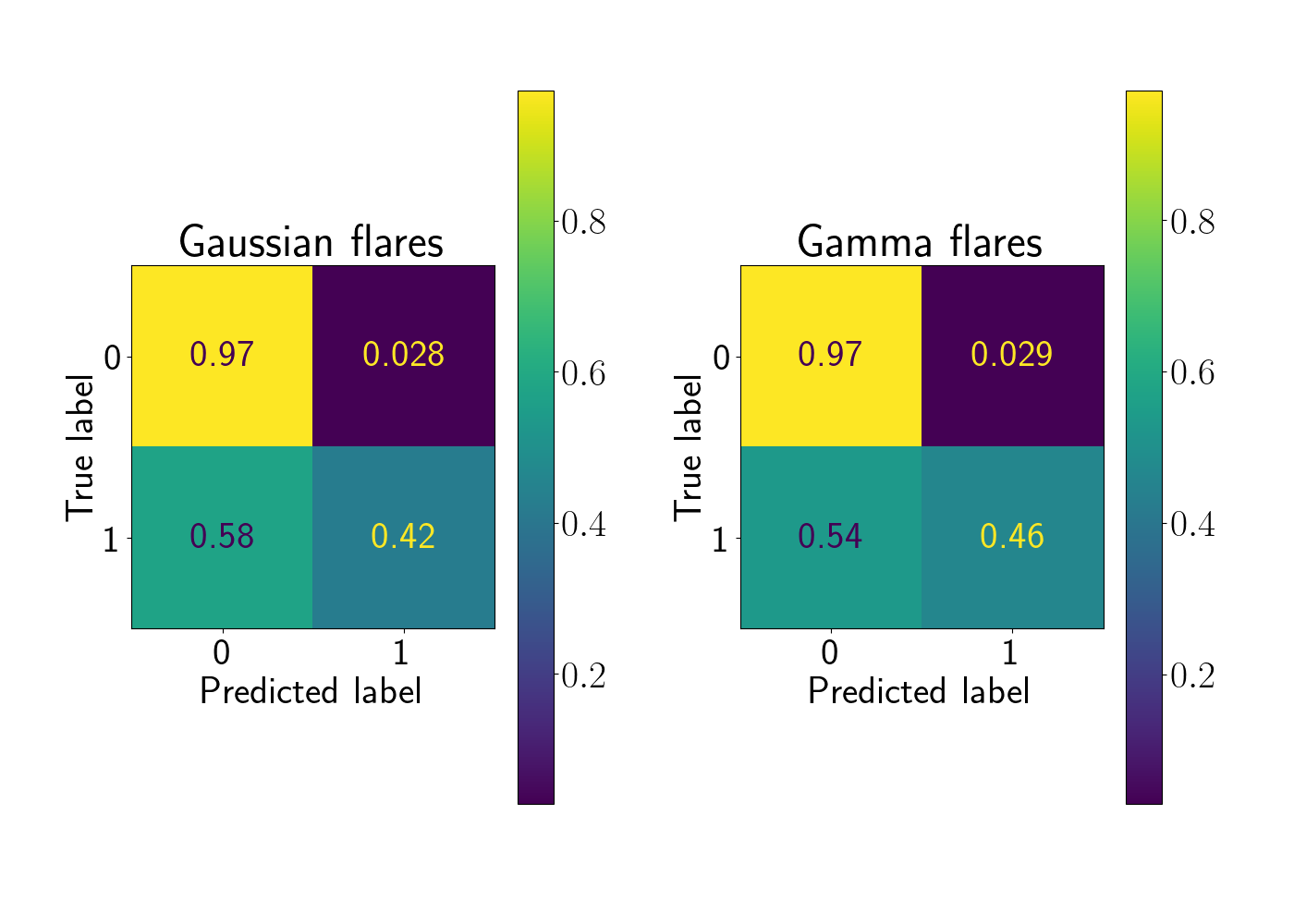}
    \caption{Confusion matrices for sub-sampled simulated light curves, in the case of injected Gaussian flares (left) and injected gamma flares (right). Note that zero and one refer to "non-flare" and "flare" respectively. The bottom right panel shows the true positive rate, which is 42 and 46 per cent for Gaussian and gamma flares respectively. The top right panel shows the false positive rate which is 2.8 per cent for Gaussian flares and 2.9 per cent for gamma flares.}
    \label{fig:subsampled_cm}
\end{figure}

As shown in Fig.~\ref{fig:subsampled_cm}, the retrieval rate reduces significantly when the light curves were sub-sampled to match ZTF cadence. There is little difference between the true positive and false positive rates of Gaussian and gamma flares, with the Gaussian flares having a true positive rate of 42 per cent and a false positive rate of 2.8 per cent. The gamma flares have a true positive rate of 46 per cent and a false positive rate of 2.9 per cent. As such, while the purity of the retrieved sample is relatively high (i.e., low false positives), the completeness is low (i.e., less than 50 per cent). 

\begin{figure}
	\includegraphics[width=\columnwidth]{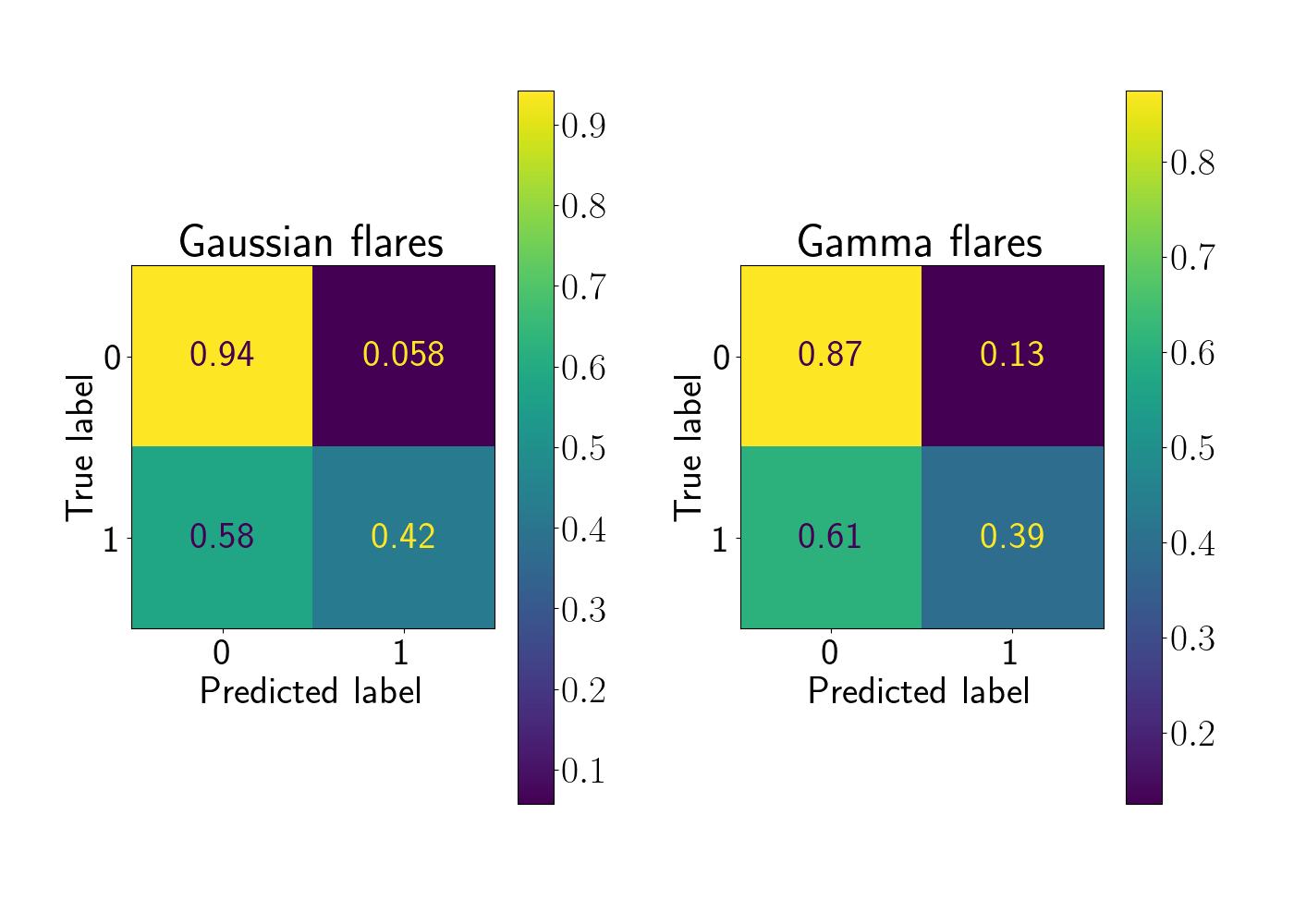}
    \caption{Confusion matrices for simulated and sub-sampled light curves with added outliers, in the case of injected Gaussian flares (left) and injected gamma flares (right). Note that zero and one refer to "non-flare" and "flare" respectively. The bottom right panel shows the true positive rate, which is 42 and 39 per cent for Gaussian and gamma flares respectively. The top right panel shows the false positive rate which is 5.8 per cent for Gaussian flares and 13 per cent for gamma flares.}
    \label{fig:outlier_cm}
\end{figure}

Fig.~\ref{fig:outlier_cm} shows the confusion matrices for simulated light curves with added outliers in the case of both Gaussian and gamma flares. This shows similar results as those found for sub-sampled light curves without outliers. In the case of Gaussian flares, the GP analysis is able to classify 94 per cent of non-flaring light curves correctly (a 6 per cent false positive rate), but only 42 per cent of flaring light curves were classified correctly. The true positive rate of the gamma flares is slightly lower at 39 per cent with a higher false positive rate of 13 per cent.

\begin{figure}
	\includegraphics[width=\columnwidth]{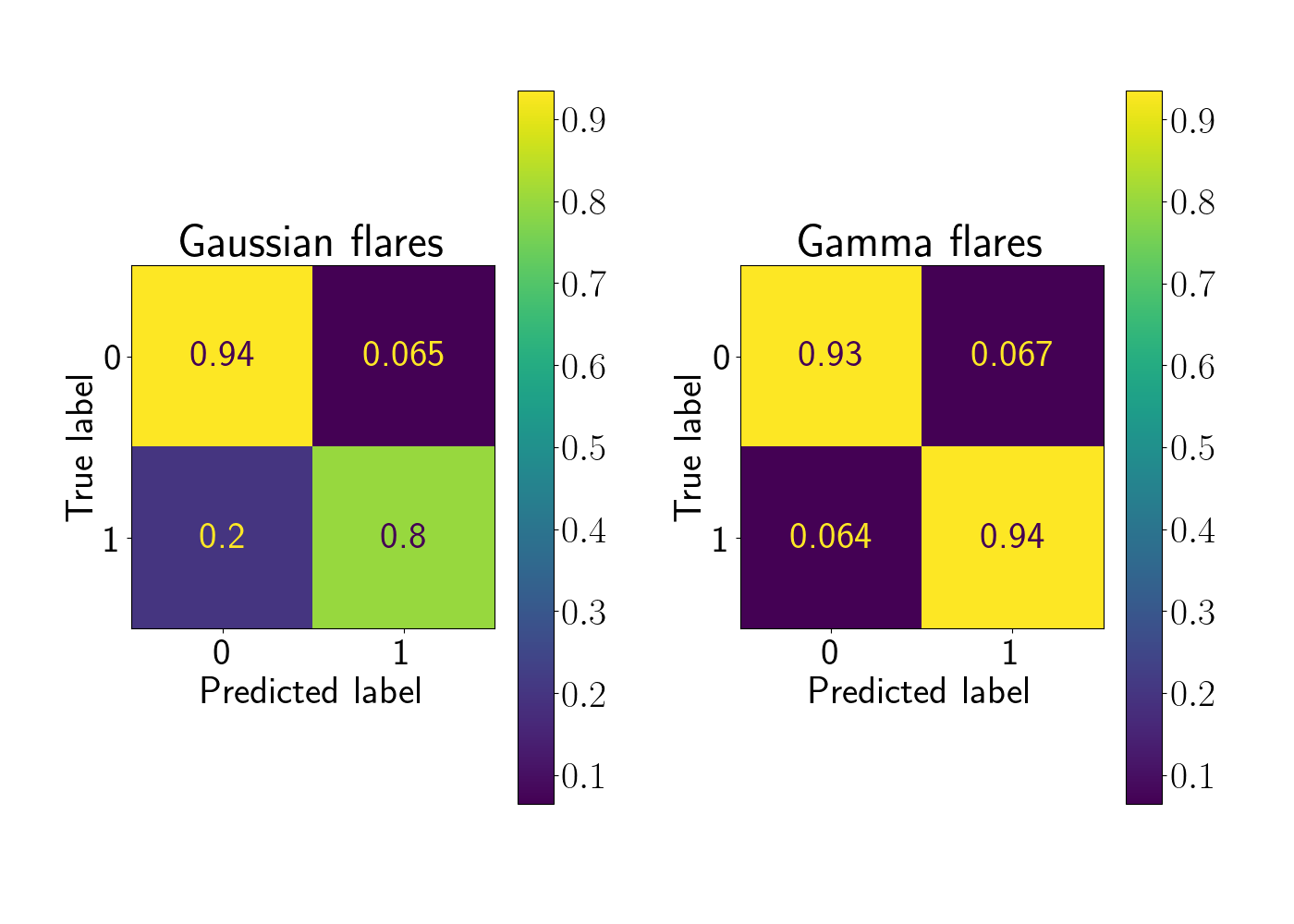}
    \caption{Confusion matrices for ZTF light curves with injected flares, in the case of injected Gaussian flares (left) and injected gamma flares (right). Note that zero and one refer to "non-flare" and "flare" respectively. The bottom right panel shows the true positive rate, which is 80 and 94 per cent for Gaussian and gamma flares respectively. The top right panel shows the false positive rate which is 6.5 per cent for Gaussian flares and 6.7 per cent for gamma flares.}
    \label{fig:ztf_cm}
\end{figure}

Despite the GP analysis struggling to identify flares in sub-sampled light curves with or without systematic outliers, we see better results with the ZTF light curves with injected flares. For these light curves, the GP analysis was more effective at classifying flares and non-flares than with the simulated sub-sampled light curves; remarkably, this is in spite of us using the ZTF cadence to sub-sample our simulated light curves. The confusion matrices for ZTF light curves with both Gaussian and gamma injected flares are shown in Fig.~\ref{fig:ztf_cm}. In the case of injected Gaussian flares, the GP analysis has an 80 per cent true positive rate and a 6.5 per cent false positive rate, compared with injected gamma flares with true and false positive rates of 94 per cent and 7 per cent respectively.

\subsection{ZTF flares}
\label{section:6.2}

The final step we took in testing the efficacy of GPs in detecting AGN flares was to perform the analysis on unadulterated AGN light curves. For this, 9035 ZTF light curves (\S\ref{section:3.5}) were analysed using a GP to determine if any would be flagged as containing flares or extreme variability.

\begin{figure}
	\includegraphics[width=\columnwidth]{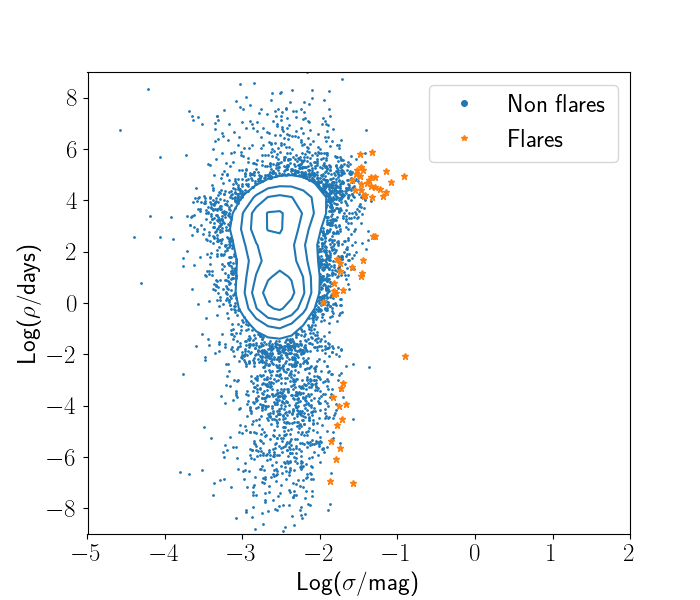}
    \caption{Distributions of hyper-parameters for real ZTF light curves of Type 1 AGN. Light curves with a posterior probability greater than 0.1, the number of data points greater than 30, the maximum spacing between consecutive data points less than 150 days, and a sigma value of greater than -2 are shown in orange. These are the resulting flare candidates.}
    \label{fig:realdata}
\end{figure}

\begin{figure}
	\includegraphics[width=\columnwidth]{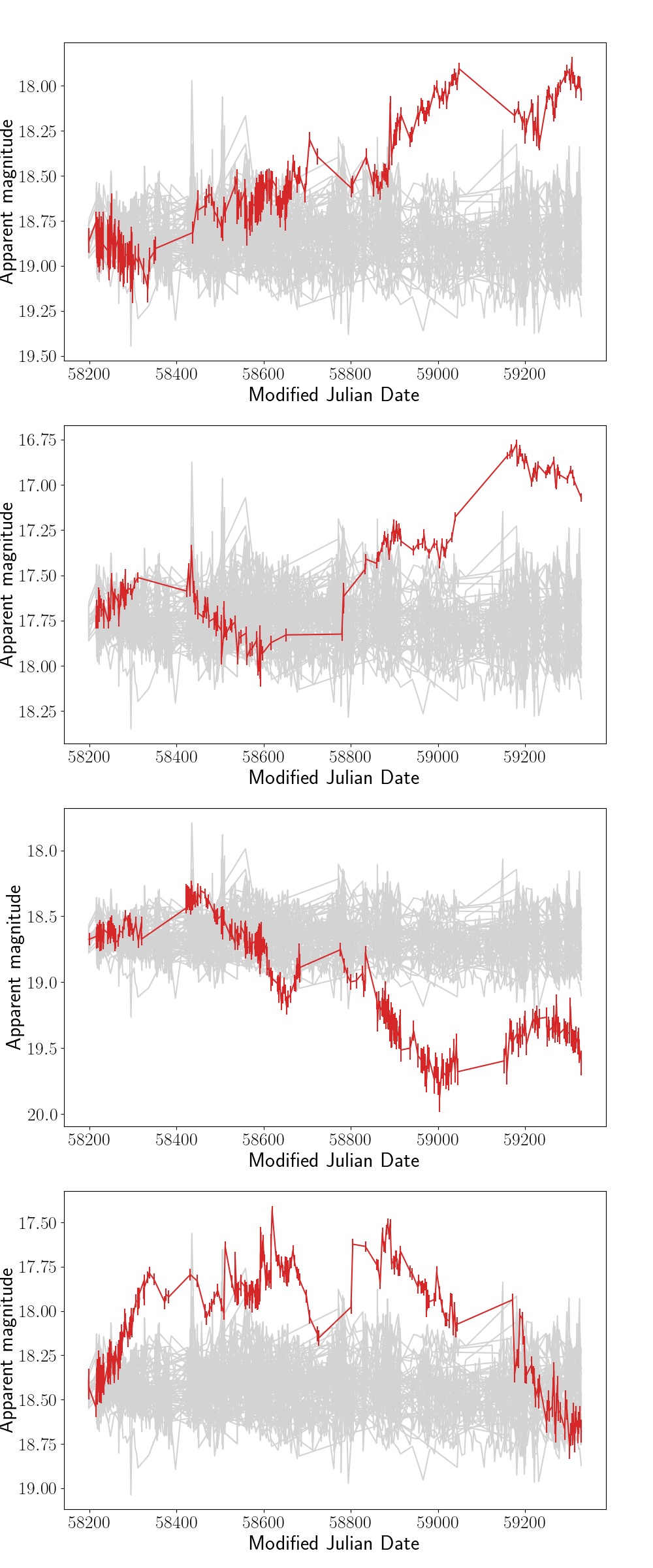}
    \caption{Four examples of ZTF light curves of flare candidates identified by the GP analysis. The red line shows the light curve of the flare candidate and the grey curves are a randomly-sampled selection of 100 light curves that were not flagged as flares by the GP analysis, demonstrating that they show extreme variability compared to the rest of the population. These light curves have been normalized for ease of visualization (see \S\ref{section:6.2}). The full sample of light curves is shown in the appendix.}
    \label{fig:ztf_lightcurves}
    \vspace{-22pt}
\end{figure}

We initially invoked a probabilistic cut-off of 0.1 for a light curve to be classified as a flare by the Gaussian Process. This cutoff resulted in a total of 257 flare candidates. On inspection, we found that a considerable number of these candidates were poorly sampled or had large gaps in their light curves. For example, 117 light curves contained fewer than 30 data points and 154 had gaps in their light curves lasting over 150 days. It is therefore feasible that some of these light curves may, indeed, contain (unsampled) flares, but we do not select them for visualisation purposes.\footnote{These, together with all our labelled light curves, are available upon request.}. It should also be noted that there were a number of light curves (117) that were assigned a high probability of containing a flare but were located in the far-left-hand side of the hyper-parameter distribution and these were removed from selection due to having low values of $\sigma$ and hence low amplitude values. Applying the above selections simultaneously and ignoring light curves with poor GP fits by visual inspection resulted in a sample of 27 flare candidates, which are shown as orange points in Fig.~\ref{fig:realdata} and whose light curves are shown in Figs. \ref{fig:ztf_lightcurves} and \ref{fig:app1}-\ref{fig:app6}.

The light curves of four examples chosen from the 27 identified flare candidates are shown in Fig.~\ref{fig:ztf_lightcurves}. In each of these four plots, we also include the light curves of 100 randomly-selected AGN that were not flagged as containing flares by the our analysis. These light curves were normalized by calculating the ``relative'' magnitude by subtracting the first magnitude value from each magnitude value in each light curve, and then adding this to the mean magnitude value of the flare light curve. Normalisation of the light curves was performed for the purpose of comparison against a common baseline and for ease of visualisation. By comparing the flare candidates to these non-flaring light curves, it is clear that the former show extreme variability. Most notably, they display longer-term, more systematic departures from their starting point relative to the comparison (non-flaring) light curves. The full sample of flare candidates is shown in the appendix. Note that our analysis is only able to detect extreme variability, and hence classifies "flares" as objects that are becoming either brighter or fainter, rather than just brighter. This is not necessarily a drawback, since if the mechanism behind AGN flares is caused by changes in accretion state, then GPs may be able to detect changing-look AGN which can both rapidly brighten or dim as their broad emission lines appear or disappear \citep[e.g.][]{LaMassa2015,Gezari2017,Yang2023}.

\section{Discussion}
\label{section:7}

In \S\ref{section:1} we described a specific problem associated with searching for flares in AGN light curves: namely, how does one detect the presence of a transient signal in data that is already intrinsically and stochastically variable? To solve this, it is necessary to quantify what constitutes a significant departure from this baseline variability in a statistical way. In this paper, we have undertaken a feasibility study to determine whether Gaussian Processes (GPs) are an effective means to achieve this. 

We find that GPs have the potential to correctly classify flaring and non-flaring simulated light curves with a high success rate, with regularly-sampled flare light curves being classified with a true positive rate of around 90 per cent for both Gaussian and gamma flares (see \S\ref{section:6.1}). However, in the case of simulated light curves that have been sub-sampled to mirror the cadence of real ZTF AGN light curves, this rate drops to around 40 to 45, depending on whether the injected flare is modelled as a Gaussian or gamma function (see \S\ref{section:6.1}). Similarly, the light curves with added outliers resulted in comparably low true positive rates (around 40 per cent; see \S\ref{section:6.1}). Despite this, when real ZTF light curves were injected with flares, the GP analysis successfully classified between 80-94 per cent per cent of the flaring light curves with false positive rates as low as 6.5 per cent (see \S\ref{section:6.2}). This false-positive rate is extremely promising, as when dealing with large amounts of data it is arguably far more important to have a low false-positive rate than a high true-positive rate, to ensure a high purity of the sample. It would be insightful to be able to place these retrieval and contamination rates in the context of other methods of finding nuclear flaring events. However, with most studies focusing -- quite reasonably -- on the identification of new flares, rather than how many they may have missed, such success rates are difficult to quantify.\footnote{The systematic comparison of different methods of finding flares is beyond the scope of this study.} Our results show that while GPs are not broadly robust against major outliers, they are still able to perform well when handling real data. It also suggests that our simulated outliers were ``pessimistic'', in that they gave the GP analysis a more difficult job than the real ZTF data. Furthermore, the retrieval rates for gamma and Gaussian flares are comparable, suggesting that the GP analysis is largely unaffected by the shape of the flare.

When we applied our GP analysis to 9035 real ZTF light curves of type 1 AGN, 27 flare candidates were identified (see \S\ref{section:6.2} and Appendix \ref{section:A1}). These light curves exhibit extreme variability when compared to 100 randomly sampled light curves that had not been flagged by the GP as flaring. 

It is tempting to take the false-negative rates of the sub-sampled flares and the ZTF injected flares to estimate the number of real flares that we could be missing. However, since those false negative rates are based on simulated data, we cannot know what the actual false negative rates are for real flares.

\begin{figure}
	\includegraphics[width=\columnwidth]{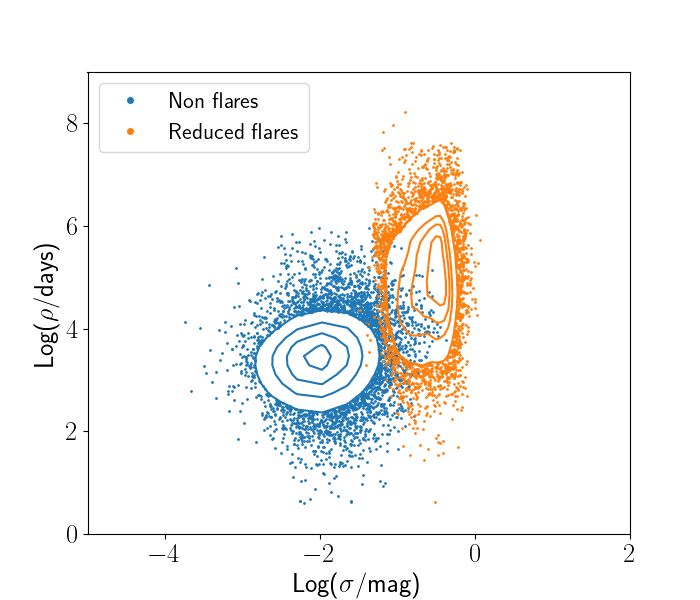}
    \caption{Distributions of hyper-parameters for sub-sampled simulated light curves that have been reduced so that the flare light curve contains only the flaring region. The separation between distributions is much greater than in Fig.~\ref{fig:Gauss_hparams}, highlighting that if one is able to localise sections of light curve it becomes more straightforward to distinguish between flares and non-flares.}
    \label{fig:chopped}
\end{figure}

We have shown that GPs are an effective way to detect extreme variability in simulated and real AGN light curves, especially in high-cadence data sets. In this paper, whilst we have demonstrated this in the r band only, it would be possible to modify the GP analysis to account for multiple bands. The use of GP analysis in light curve classification is not without caveats, however, as there are a number of limitations. As we have shown, a GP is not robust against extreme outliers. In addition, GPs optimise the kernel hyper-parameters across the whole light curve which favours the detection of longer-duration, larger-amplitude flares (and especially those that span a significant fraction of the light curve). To investigate the impact of these ``average'' hyper-parameters, we sliced the sub-sampled, simulated flare light curves so that they contained only the flaring region of the light curve and repeated our analysis. The results are shown in Fig.~\ref{fig:chopped}. This shows that if a GP is somehow able to simultaneously ``focus'' on subsections of a light curve it would have a much higher success rate in terms of distinguishing between flares and non-flares. This demonstrates that GPs could be even more effective at flare classification if it were able to calculate a light curve's hyper-parameters in a more localised way.

Furthermore, whilst they can quantify the probability of a light curve containing a flare, the GP analysis performed here cannot specify the location of the flare within the light curve. This is not necessarily a pitfall when searching for flares or extreme variability in archival data, but in the era of time-domain astronomy where surveys such as the LSST will detect potentially millions of transient sources per night, it warrants the ability to detect an AGN flare in real-time, ideally before it peaks. Such a requirement clearly demands an alteration of this method to enable the detection of flares as they happen.

These limitations highlighted above motivate the need to build on our techniques with the intention of localising flares within light curves. Two possible ways of achieving this are: tracking the posterior flare probability, $P(\sigma, \rho, \theta | y)$ as a function of time whilst feeding the GP new data, or calculating the posterior probability $P(y_{new}|y_{data})$ to determine whether new points in a light curve can be described by the current GP regime and flag them as a flare otherwise. However, these methods are potentially computationally intensive and so it is important to be able to devise a means of flare localisation in an efficient way. This may require more sophisticated techniques such as deep Gaussian Processes \citep[]{Damianou2013} where the choice of kernel function will depend on training data. Other possibilities include change-point detection \citep[]{Graham2023} or regime-switching models \citep[]{Hamilton2010}. Investigation into these more complex GP techniques is, however, beyond the scope of this study.

\section{Summary}
\label{section:8}

We have undertaken a feasibility study to investigate whether Gaussian Processes (GPs) are an effective means of identifying and classifying AGN flares in optical light curves. Using a combination of simulated and real AGN light curves, we used GP analysis to investigate how the distributions of kernel hyper-parameters change after the injection of a simulated AGN flare into a light curve (\S\ref{section:4}). We then used these distributions as a basis to classify light curves in terms of whether they contain a flare or not, and calculate corresponding flare retrieval rates (\S\ref{section:5}). Throughout, we exploited five different classes of light curve, each more representative of real light curves than the last: 

\begin{enumerate}
  \item injecting flares and simulated with a constant 10-day cadence;
  \item injecting flares and sub-sampled to match the cadence of real ZTF light curves;
  \item as (ii), but with added outliers;
  \item real ZTF with injected flares;
  \item real ZTF light curves.
\end{enumerate}

In the case of (i), we find that the kernel hyper-parameter distributions for flares and non-flares exist in different but partially overlapping regions of parameter space (\S\ref{section:4}). This means that whilst the distributions can never be separated completely, because GPs are statistically robust, GP analysis can be used to distinguish between the distributions in a probabilistic way. In the cases of light curve classes (ii)-(v), however, the hyper-parameter distributions for flares and non-flares overlap significantly more than in the case of (i). Despite this, we are able to demonstrate that GPs can be used as a classification tool for AGN flares, with varying degrees of success. Our results can be summarised as follows:

\begin{enumerate}
  \item For simulated flares with a ten-day cadence and with injected flares, we find a true positive rate of 91-92 per cent and a false positive rate of 7-11 per cent for Gaussian and gamma flares respectively.
  \item When the light curves in (i) are sub-sampled to match the cadence of our sample of ZTF light curves (\S\ref{section:3.5}), the true positive rates reduce significantly to 42-46 per cent for Gaussian and gamma flares respectively, though the false positive rate is found to be approximately 3 per cent in each case.
  \item When outliers are added to these simulated light curves, the true positive rates remain similar to those found in (ii), although the false positive rates increase to 6 and 13 per cent for Gaussian and gamma flares, respectively.
  \item When our sample of real AGN light curves are injected with simulated Gaussian and gamma flares, the results are more promising than in the cases of (ii-iii). We obtain true positive rates of 80 and 94 per cent for Gaussian and gamma flares, respectively, while the false positive rates remain similar as to that found for class (iii) at approximately 7 per cent.
  \item Finally, we applied our GP analysis to the unadulterated sample of ZTF light curves to determine whether any real AGN light curves would be flagged as containing flares by our GP analysis. As shown in \S\ref{section:6.2}, the GP analysis classified 27 out of 9035 AGN light curves as containing flares or extreme variability. When compared with a randomly-selected sample of 100 light curves that were not flagged as flares, they indeed show greater levels of variability, particularly in the form of longer-term, systemic departures from their starting point.
\end{enumerate}

Overall, we have demonstrated that GP analysis can be used to calculate the probability that an incoming AGN light curve contains a flare. We find that this is a promising method to detect flares in otherwise variable optical light curves, although it can be negatively affected by extreme outliers and poorly sampled data. In order to keep up with the large amounts of data involved in future surveys such as the LSST \citep[]{Ivezi2019}, there is a growing requirement to be able to detect AGN flares and transients alike before they peak to enable for rapid follow-up. Therefore, since our GP analysis in this work is able to calculate the probability of an incoming light curve containing a flare but not the exact location of the flare within the light curve, there is a need to build on this GP technique to be able to localise a flare as it happens. As mentioned in \S\ref{section:7}, these techniques may be computationally intensive and so further feasibility studies are required to determine the most efficient way to achieve flare localisation within a light curve.

\section*{Acknowledgements}

SAJM is supported by an STFC-funded PhD studentship. JRM is supported by STFC funding for UK participation in LSST, through grant ST/Y00292X/1. SPL is funded by STFC through grant ST/V000853/1. 

SAJM thanks Professor P. Crowther for his helpful feedback. The authors thank Dr. N. Caplar for his constructive comments.

This work made use of Astropy:\footnote{http://www.astropy.org} a community-developed core Python package and an ecosystem of tools and resources for astronomy \citep{astropy:2013, astropy:2018, astropy:2022}. This work is based on observations obtained with the Samuel Oschin Telescope 48-inch and the 60-inch Telescope at the Palomar Observatory as part of the Zwicky Transient Facility project. ZTF is supported by the National Science Foundation under Grant No. AST-2034437 and a collaboration including Caltech, IPAC, the Weizmann Institute for Science, the Oskar Klein Center at Stockholm University, the University of Maryland, Deutsches Elektronen-Synchrotron and Humboldt University, the TANGO Consortium of Taiwan, the University of Wisconsin at Milwaukee, Trinity College Dublin, Lawrence Livermore National Laboratories, and IN2P3, France. Operations are conducted by COO, IPAC, and UW. 

\section*{Data Availability}

The data used in this work can be accessed by reasonable request to the corresponding author.



\bibliographystyle{mnras}
\bibliography{example} 



\appendix

\section{ZTF light curves showing extreme variability}
\label{section:A1}

In this section we present the ZTF light curves of AGN that were classified as flares by the GP. Note that some objects are actually decreasing in brightness as the GP detects extreme variability in both directions.

\begin{figure}
	\includegraphics[width=\columnwidth]{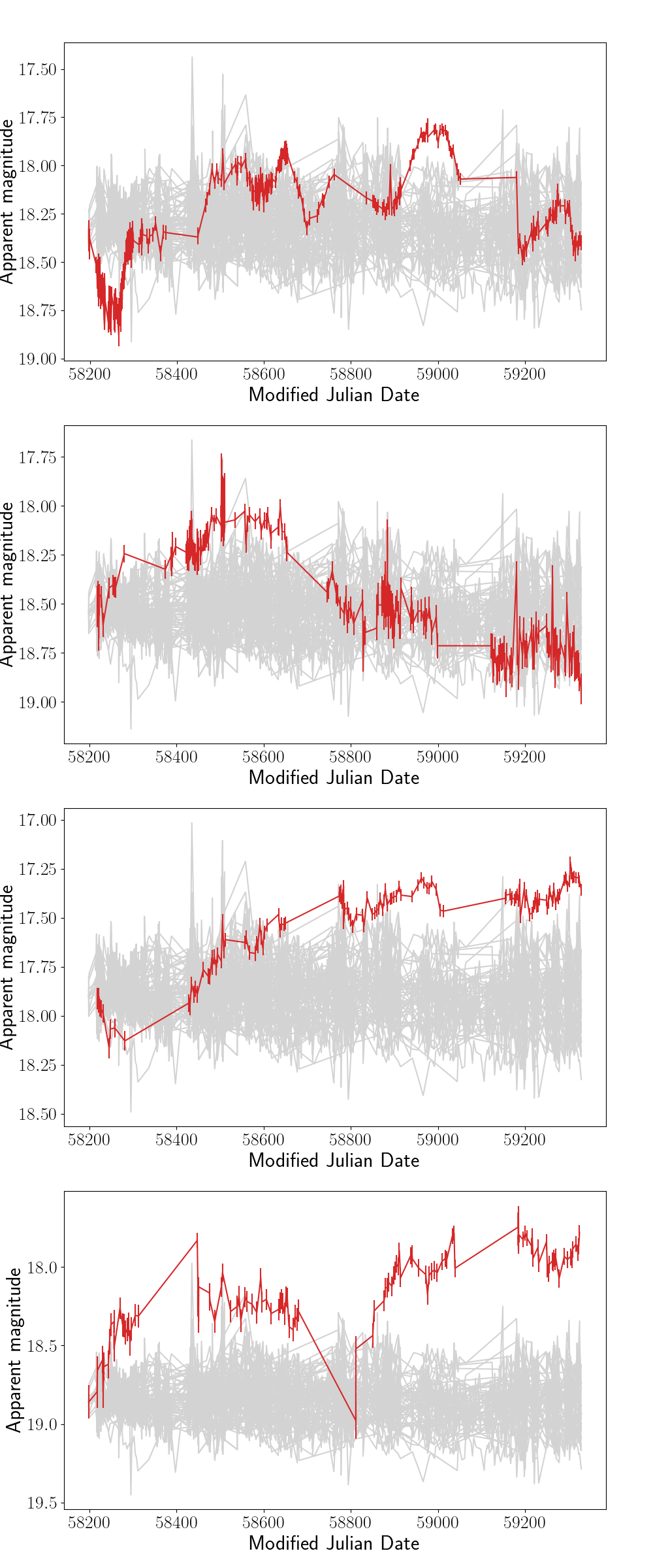}
    \caption{ZTF light curves of flare candidates identified by the GP. The red line shows the light curve of the flare candidate and the grey curves are a randomly-sampled selection of 100 light curves that were not flagged as flares by the GP, demonstrating that they show extreme variability compared to the rest of the population. These light curves have been normalized for ease of visualization (see \S\ref{section:6.2}).}
    \label{fig:app1}
    \vspace{-2pt}
\end{figure}

\begin{figure}
	\includegraphics[width=\columnwidth]{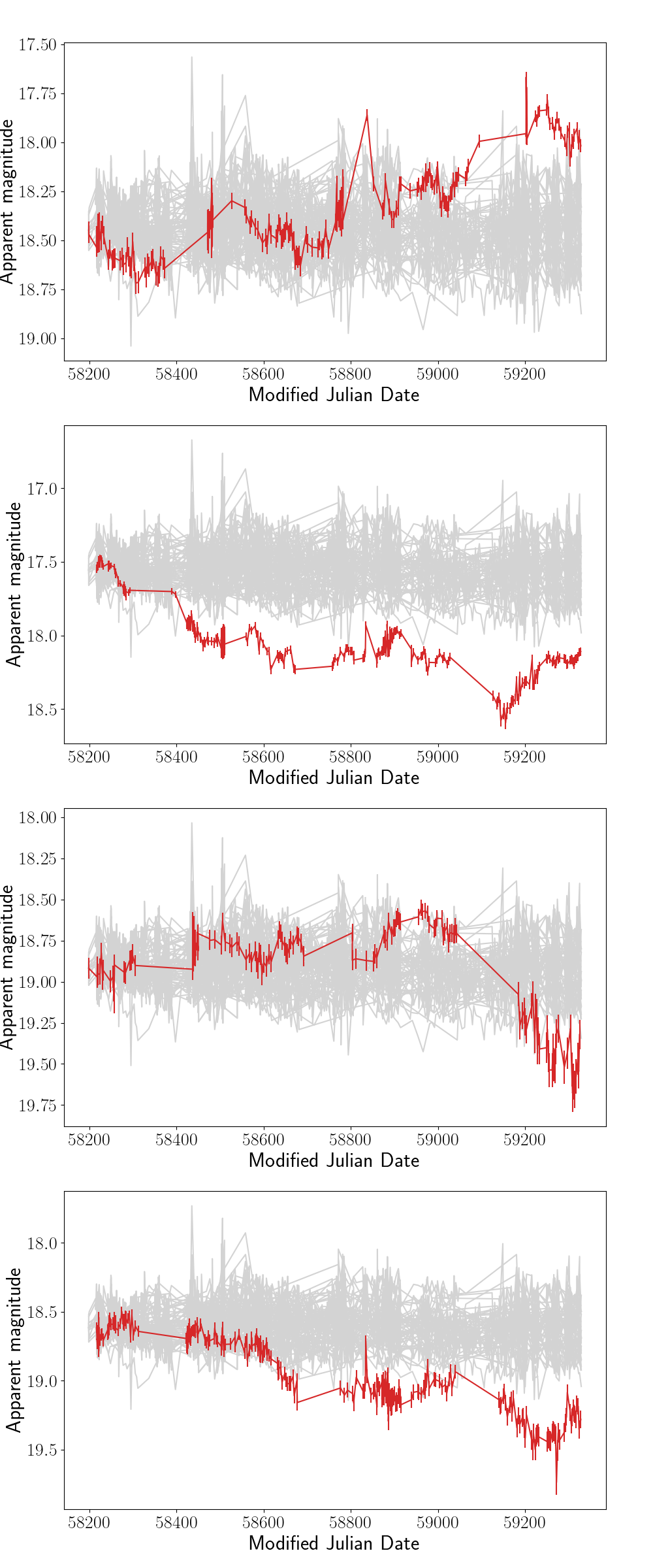}
    \caption{ZTF light curves of flare candidates identified by the GP. The red line shows the light curve of the flare candidate and the grey curves are a randomly-sampled selection of 100 light curves that were not flagged as flares by the GP, demonstrating that they show extreme variability compared to the rest of the population. These light curves have been normalized for ease of visualization (see \S\ref{section:6.2}).}
    \label{fig:app2}
    \vspace{-2pt}
\end{figure}

\begin{figure}
	\includegraphics[width=\columnwidth]{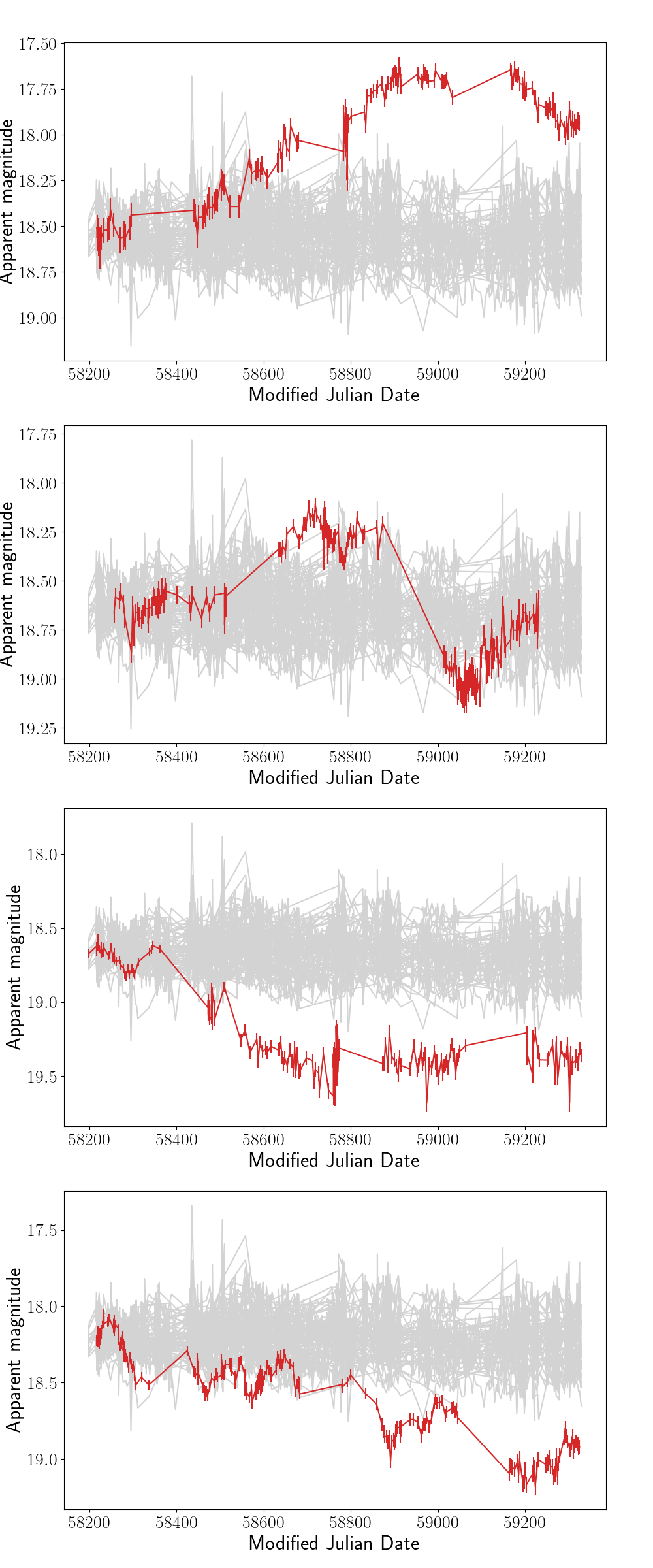}
    \caption{ZTF light curves of flare candidates identified by the GP. The red line shows the light curve of the flare candidate and the grey curves are a randomly-sampled selection of 100 light curves that were not flagged as flares by the GP, demonstrating that they show extreme variability compared to the rest of the population. These light curves have been normalized for ease of visualization (see \S\ref{section:6.2}).}
    \label{fig:app3}
    \vspace{-2pt}
\end{figure}

\begin{figure}
	\includegraphics[width=\columnwidth]{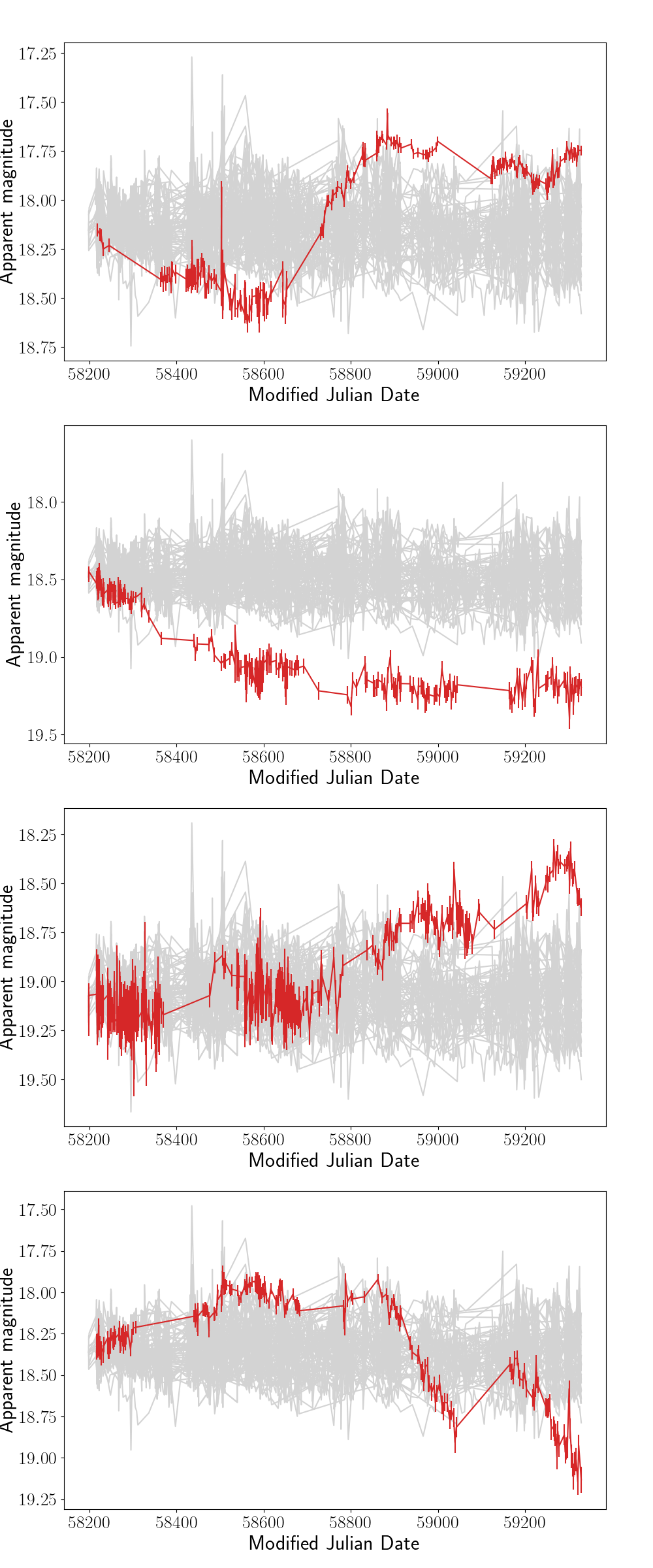}
    \caption{ZTF light curves of flare candidates identified by the GP. The red line shows the light curve of the flare candidate and the grey curves are a randomly-sampled selection of 100 light curves that were not flagged as flares by the GP, demonstrating that they show extreme variability compared to the rest of the population. These light curves have been normalized for ease of visualization (see \S\ref{section:6.2}).}
    \label{fig:app4}
    \vspace{-2pt}
\end{figure}

\begin{figure}
	\includegraphics[width=\columnwidth]{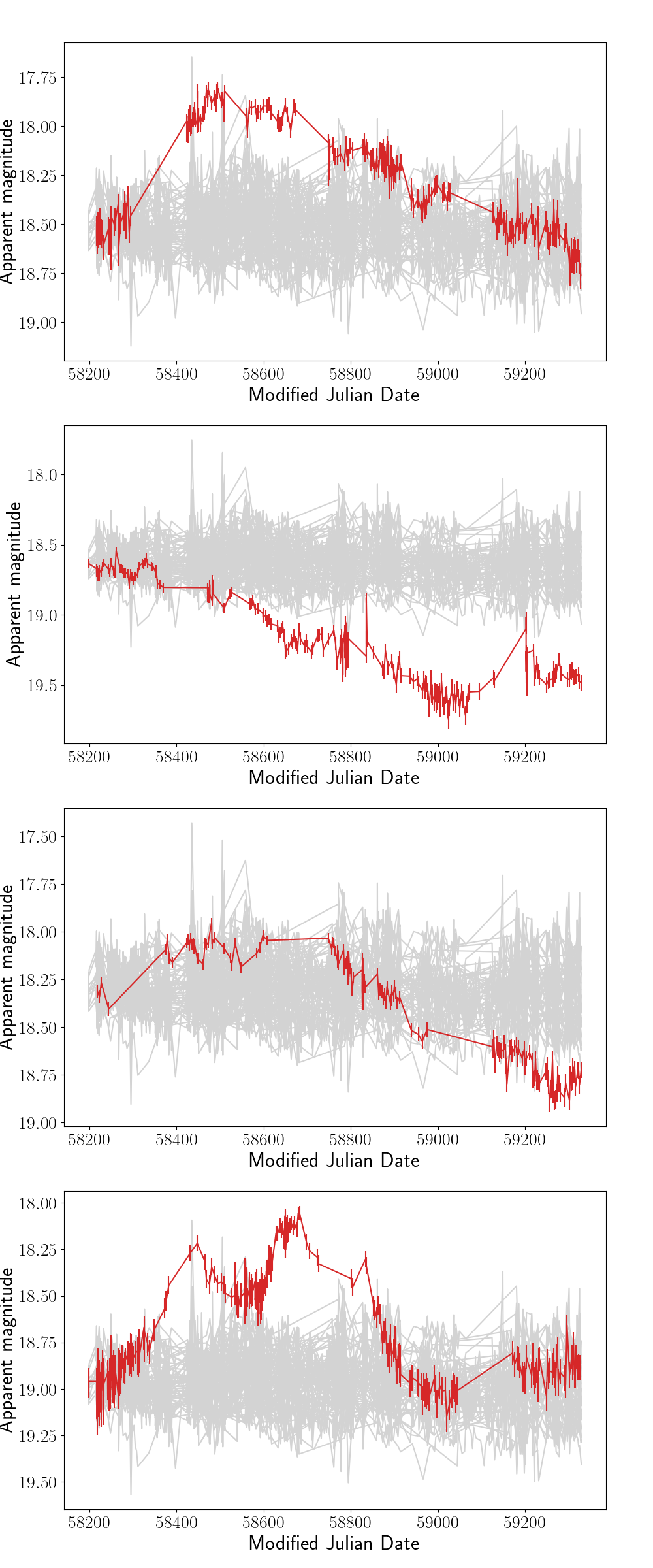}
    \caption{ZTF light curves of flare candidates identified by the GP. The red line shows the light curve of the flare candidate and the grey curves are a randomly-sampled selection of 100 light curves that were not flagged as flares by the GP, demonstrating that they show extreme variability compared to the rest of the population. These light curves have been normalized for ease of visualization (see \S\ref{section:6.2}).}
    \label{fig:app5}
    \vspace{-2pt}
\end{figure}

\begin{figure}
	\includegraphics[width=\columnwidth]{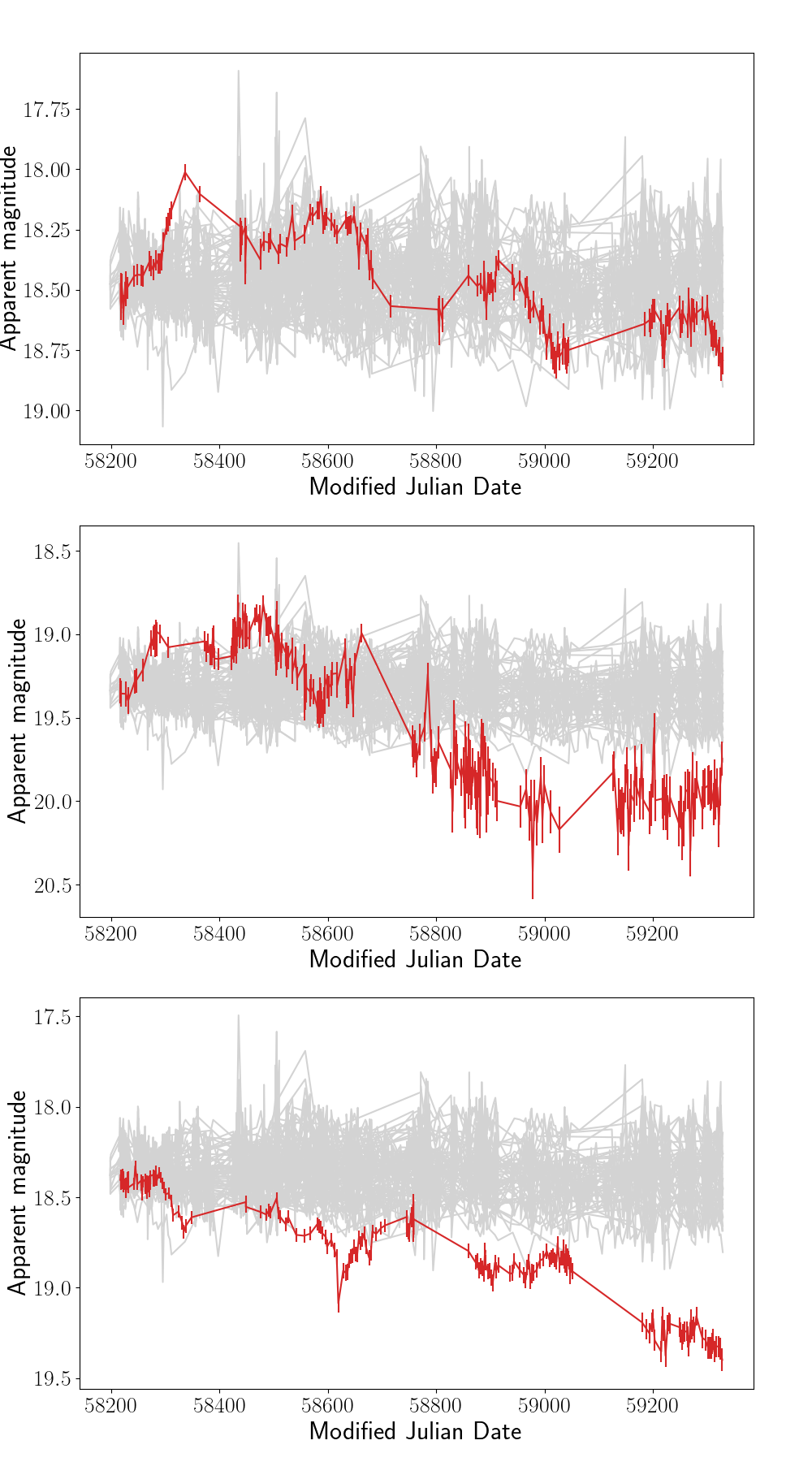}
    \caption{ZTF light curves of flare candidates identified by the GP. The red line shows the light curve of the flare candidate and the grey curves are a randomly-sampled selection of 100 light curves that were not flagged as flares by the GP, demonstrating that they show extreme variability compared to the rest of the population. These light curves have been normalized for ease of visualization (see \S\ref{section:6.2}).}
    \label{fig:app6}
    \vspace{-2pt}
\end{figure}


\bsp	
\label{lastpage}
\end{document}